\documentclass[aps,superscriptaddress,showkeys,nofootinbib]{revtex4}

\usepackage[dvips]{color}                 
\usepackage{eucal}

\usepackage{float}
\usepackage{ulem}

\usepackage{epsfig}
\usepackage{xcolor}

\begin{document}

\title{Tricritical behavior of a relativistic field theory in one dimension}

\author{Heron Caldas, A. L. Mota}

\affiliation
{Departamento de
  Ci\^{e}ncias Naturais. Universidade Federal de S\~{a}o Jo\~{a}o del Rei,\\
  36301-160, S\~{a}o Jo\~{a}o del Rei, MG, Brazil}

\begin{abstract}

The tricritical behavior in a class of one-dimensional (1D) field theories that exhibit spontaneous symmetry breaking at zero temperature and chemical potential is analyzed. 
In the Gross-Neveu (GN)-type models of massless fermions the discrete chiral symmetry is spontaneously broken. After doping, the symmetry is restored at a critical chemical potential. We investigate the temperature effects on this doped 1D system under an external constant Zeeman magnetic field $B_0$. We find that $B_0$ suppresses the gapless behavior present for certain values of chemical potential and is able to induce a gapless-gapped phase transition at a critical field strength. We also discuss about the consequences of the consideration of inhomogeneous condensates to the tricritical point, within the Ginzburg-Landau expansion.

\end{abstract}



\maketitle



\section{Introduction}

The Gross-Neveu model in $1+1$ dimension (1D) is a Quantum Field Theory (QFT) with $N$ flavors of massless fermions with a four-fermion interaction term~\cite{GN}. The GN model in 1D has been used not only as a theoretical model which displays some basic features of two flavor massless QCD: $1/N$ expansion, renormalization, asymptotic freedom, chiral symmetry breakdown (in the vacuum), and dimensional transmutation~\cite{Coleman}. The GN model has also been employed in the study of other effects in QFT, such as, for example, the pion condensation phenomenon~\cite{Klimenko1} and magnetic catalysis~\cite{Klimenko2,Klimenko3,Lenz1}. In condensed matter physics, for instance, the GN model serves to describe the phase transition from a solitonic to a metallic phase in {\it trans}-polyacetylene (TPA)~\cite{CM}, and also as an effective field theory for TPA~\cite{CaldasJSTAT}. Thus, due to the simplicity and ``moldability'' of the GN model to approach different problems from distinct areas of physics, there has been recently a renewed interest in its use in the investigation of these aforementioned and other related topics~\cite{Thies1,Lenz1,Lenz2,Lenz3,Lobos,Nei,Thies2,Thies3}.

In the field theory language, the discrete chiral symmetry is spontaneously broken (at zero temperature and chemical potential) in the GN model in 1D, and there is a dynamical generation of a mass (gap) for the electrons. The global external chemical potential $\mu$ is introduced in the theory to represent the extra electrons that are inserted in the system by a doping process. As is well known, at zero temperature the GN model undergoes a first order phase transition to a symmetry restored (zero gap) phase at a critical chemical potential $\mu_c=\frac{\Delta_0}{\sqrt{2}}$, where $\Delta_0$ is the constant band-gap. At zero or small chemical potential ($\mu \ll T$), the system undergoes a second-order phase transition to the unbroken symmetry phase as the temperature is increased~\cite{Wolff,Mao,Chodos}. Along the line separating the gapped and gapless phases, there is a tricritical point~\cite{TCP}. These phase transitions are exhibited in the phase diagrams that we discuss in detail below.

The study of the effects of an external Zeeman magnetic field $B_0$ applied on a system is important in many physical situations, as, for instance, in the investigation of metal-insulator transition and magnetization in 1D~\cite{Caldas2} and 2D~\cite{Dent,PRB2} electron systems. We investigate the consequences of a $B_0$ magnetic field applied parallel to the doped 1D system at zero and finite T. This produces an imbalance $\delta\mu \propto B_0$ between the chemical potential of the spin-$\uparrow$ and spin-$\downarrow$ electrons added in the system. We show that with low imbalance (with respect to the critical imbalance $\delta\mu_c$), the system acquires a partial spin polarization. With a strong enough imbalance, there is a total absence of electrons from one of the two possible spin orientations ($\uparrow$ or $\downarrow$), meaning that the system is now fully polarized. We expect that this might have observable consequences in the transport and magnetic properties of one-dimensional organic conductors, such as doped TPA. We also consider spatially inhomogeneous gap parameter $\Delta(x)$ and obtain a $x$-dependent grand potential density $V_{eff}(x)$.

We now summarize our main results for a one-dimensional fermion system described by the model in Sec.~\ref{GN-ModelH}.
\newline
(i) We show that at zero temperature and finite chemical potential, an applied Zeeman field $B_0$ suppresses the gapless phase present for certain values of chemical potential and is able to drive a gapless-gapped transition at a critical field strength $B_{0,c}$.
\newline
(ii) This critical field $B_{0,c}$ is also responsible for the fully polarization of the system, which therefore results in a noninteracting system.
\newline
(iii) After performing a Ginzburg-Landau expansion of the effective potential up to orders $(\Delta/k_BT)^6$ and $(\mu_{\uparrow,\downarrow}/k_BT)^4$, the phase diagrams in the plane $(\mu_{\uparrow,\downarrow},T)$ are obtained. This allows the identification of the first and second-order lines, as well as the tricritical point.
\newline
(iv) The role of the external Zeeman field $B_0$ is to move the tricritical point in the opposite direction to the growth of $B_0$. This means that the gapped region of the phase diagram is significantly reduced.
\newline
(v) We find that at zero temperature, and for $\mu < \mu_c$, there is a ``reentrant'' behavior as we increase $\delta \mu$ (i.e., $B_0$), with gapped-gapless-gapped phases.
\newline
(vi) The gapless-gapped phase transition at a critical field strength $B_{0,c}$ as a function of the temperature is also investigated. The reentrant phenomena is not observed as the temperature increases.
\newline
(vii) The location of the tricritical point of the GN phase diagram under the influence of a given external Zeeman magnetic field is unaltered even considering spatially inhomogeneous $\Delta(x)$ condensates.

The rest of the paper is organized as follows. In Section~\ref{GN-ModelH}, we present the Gross-Neveu model Lagrangian and some basic definitions. In Section~\ref{rep}, we obtain the renormalized effective potential at zero and finite chemical potential and temperature, $V_{eff}$. With $V_{eff}$ in hand, we construct several phase diagrams of our model and discuss various phase transitions that can be obtained by varying model parameters such as chemical potential, temperature, and magnetic field. In this section, we also briefly discuss some relevant magnetic properties of this system. In Section~\ref{IC} we consider a spatially inhomogeneous gap parameter $\Delta(x)$ and discuss interesting (particular) limits of the averaged grand potential density $V_{eff}(x)$. Finally, in Section~\ref{Summary} we present our conclusions and summarize the main results.

\section{The Gross-Neveu Model}
\label{GN-ModelH}

\subsection{The Gross-Neveu Model Lagrangian}

The Lagrangian density of the massless GN model~\cite{GN} reads

\begin{eqnarray}
{\cal L}_{\rm GN} =
\sum_{s=\uparrow,\downarrow} \bar \psi^{s} \left( i \hbar \gamma_0 \partial_t - i \hbar v_F \gamma_1 \partial_x \right) \psi^s  
+ \frac{  \lambda \hbar v_F}{2 N} \left( \bar\psi \psi  \right)^2 \;,
\label{LagGN}
\end{eqnarray}
where $\psi^s$ is a Dirac fermion with $N$ flavors, $s$ is an internal symmetry index (spin) determining the effective degeneracy of the fermions, $v_F$ is the Fermi velocity and $\lambda$ is the bare coupling constant. The fermion bilinears have sums over flavor indices $\bar\psi \psi = \bar\psi^j \psi^j = \sum_{j=1}^{N} \bar\psi^j \psi^j$. The Dirac $\gamma$ matrices are given, in terms of the Pauli matrices, by

$
\gamma^0 = \sigma_x = \left[ \begin{array}{cc}
0 & 1 \\ 
1 & 0
\end{array} \right],
\gamma^1 = i \sigma_y = \left[ \begin{array}{cc}
0 & 1 \\ 
-1 & 0
\end{array} \right], \label{DiracMatrices}
$
and $\gamma^5 = \gamma^0 \gamma^1$. In the large-{N} (mean-field) approximation, one obtains~\cite{GN,Wolff,PRB}

\begin{eqnarray}
{\cal L}_{\rm GN} =
\sum_s {\psi^s}^{\dagger}
\left( i  \hbar \partial_t - i \hbar v_F \gamma_5 \partial_x - \gamma_0
\Delta(x) 
\right) \psi^s 
-\frac{N}{2  \hbar v_F \lambda} 
\Delta^2(x),
\label{LagGN2}
\end{eqnarray}
where we have used that $\bar \psi \equiv {\psi}^\dagger \gamma_0$, and $\Delta(x)$ is the spatially inhomogeneous gap (condensate). A chemical potential $\mu$ is added to the above Lagrangian as 

\begin{eqnarray}
{\cal L}_{\rm CP} = \sum_{s=\uparrow,\downarrow} \mu \bar \psi^s \gamma_0 \psi^s.
\label{LagCP}
\end{eqnarray}

In order to consider the application of an external parallel Zeeman magnetic field (which couples to the electrons' spin) to the system and its effects, a Zeeman energy term 

\begin{eqnarray}
{\cal L}_{\rm Z} = \sum_{s=\uparrow,\downarrow} \frac{\sigma_s}{2} g \mu_B B_0 \bar \psi^s \gamma_0 \psi^s,
\label{LagZ}
\end{eqnarray}
where $\sigma_\uparrow= + 1$ and $\sigma_\downarrow= - 1$ is also included in ${\cal L}_{\rm GN}$. Then, it is convenient to start by writing the grand canonical partition function associated with ${\cal L}_{\rm GN}$ in the imaginary time formalism~\cite{Kapusta}

\begin{equation}
\label{action1}
{\cal Z}= \int  ~ {\cal D} \bar \psi ~ {\cal D} \psi~ exp \left\{ \int_0^{ \beta} d\tau \int dx~ \left[ { L}_{\rm GN} \right] \right\},
\end{equation}
where $\beta=1/k_BT$, $k_B$ is the Boltzmann constant, and $L_{\rm GN}$ is the Euclidean GN Lagrangian density:

\begin{equation}
\label{L1}
{ L}_{\rm GN}= \sum_{s=\uparrow,\downarrow} \bar \psi^s [- \gamma_0 \hbar \partial_\tau + i \hbar v_F \gamma_1 \partial_x -  \Delta(x) +  \gamma_0 \mu_s  ] \psi^s -\frac{1}{2 \hbar v_F  \lambda}  \Delta^2(x),
\end{equation}
where we have defined effective chemical potentials as $\mu_\uparrow=  \mu + \delta \mu$, $\mu_\downarrow= \mu - \delta \mu$. Thus, we have seen that te application of a static magnetic field on the system results in a Zeeman splitting energy given by $\Delta E= S_z g \mu_B B_0$~\cite{Madelung,Kittel}, where $S_z=\pm 1/2$, $g$ is the effective $g$-factor and $\mu_B=e \hbar/2m \approx  5.788 \times 10^{-5} ~{\rm eV}~ {\rm T}^{-1}$ is the Bohr magneton, $m$ is the bare electron mass, and $B_0$ is the magnetic field strength, giving $\delta \mu = \frac{g}{2} \mu_B B_0$. 

Integrating over the fermion fields leads to 

\begin{equation}
\label{action2}
{\cal Z} =  exp~{ \left\{ -\frac{\beta}{ 2 \hbar v_F  \lambda} \int dx ~ \Delta^2(x) \right\}} ~\Pi_{s=\uparrow,\downarrow} det D_s,
\end{equation}
where $D_s=- \gamma_0 \partial_\tau + i \hbar v_F \gamma_1 \partial_x +  \gamma_0 \mu_s -  \Delta(x) $ is the Dirac operator at finite temperature and density. Since $\Delta(x)$ is static, one can transform $D_s$ to the $\omega_n$ plane, where $\omega_n=(2n+1)\pi T$ are the Matsubara frequencies for fermions, yielding $D_s= (- i \omega_n + \mu_s)\gamma_0 + i \hbar v_F \gamma_1 \partial_x  -  \Delta(x) $. After using an elementary identity $\ln (det(D_s))={\rm Tr} \ln(D_s)$, one can define the bare effective action for the static $\Delta(x)$ condensate

\begin{equation}
\label{action3}
S_{eff} [\Delta]= -\frac{\beta}{ 2 \hbar v_F  \lambda}  \int dx ~ \Delta^2(x) + \sum_{s=\uparrow,\downarrow} {{\rm Tr} \ln (D_s)},
\end{equation}
where the trace is to be taken over both Dirac and functional indices. The condition to find the stationary points of $S_{eff} [\Delta]$ reads

\begin{equation}
\label{action4}
\frac{\delta S_{eff} [\Delta]}{\delta \Delta(x)}=0= -\frac{ \beta}{  \hbar v_F  \lambda} \Delta(x) + \frac{\delta}{\delta \Delta(x)} \left[ \sum_{s=\uparrow,\downarrow} {{\rm Tr} \ln (D_s)} \right].
\end{equation}
The equation above is a complicated and generally unknown functional equation for $\Delta(x)$, whose solution has been investigated several times in the literature~\cite{Dashen,Feinberg,gnpolymers,gnpolymers1,gnpolymers2,gnpolymers3,gnpolymers4,Basar,Lenz2}. Its solution is not only of academic interest but has direct application in condensed matter physics as, for example, in~\cite{SSH,Review,gnpolymers3}, and in the present work.

\section{The Renormalized Effective Potential at Zero and Finite Temperature}
\label{rep}

For a constant $\Delta$ field the Dirac operator reads $D_s= (- i \omega_n + \mu_s)\gamma_0 + i \hbar v_F \gamma_1 p  -  \Delta $, so the trace in Eq.~(\ref{action3}) can be evaluated in a closed form for the asymmetrical ($\delta \mu \neq 0$) system~\cite{Caldas2}. From Eq.~(\ref{action2}) one obtains the {\it effective potential} $V_{eff}=-\frac{k_B T}{L} \ln {\cal Z}$, where $L$ is the length of the system

\begin{eqnarray}
\label{poteff}
V_{eff}(\Delta,\mu_{\uparrow,\downarrow},T)=\frac{1}{ 2 \hbar v_F  \lambda} \Delta^2 
&-& k_B T \int^{+\infty}_{-\infty}{\frac{dp}{4 \pi \hbar}}~ \Big[ 2 \beta E_p
+  \ln \left(1+e^{-\beta E_\uparrow^+}\right)+  \ln \left(1+e^{-\beta E_\uparrow^-}\right)\\
\nonumber
&+& \ln \left(1+e^{-\beta E_\downarrow^+}\right) + \ln \left(1+e^{-\beta E_\downarrow^-}\right) \Big],
\end{eqnarray}
where $E_{\uparrow,\downarrow}^{\pm} \equiv E_p \pm \mu_{\uparrow,\downarrow}$, $E_p = \sqrt{v_F^2 p^2+\Delta^2}$.

\subsubsection{Zero Temperature and Zero Chemical Potentials}

The first term in the integration in $p$ in Eq.~(\ref{poteff}), corresponding to the vacuum part ($\mu_{\uparrow,\downarrow}=T=0$), is divergent. Introducing a momentum cutoff $\Lambda$ to regulate this part of $V_{eff}$, one obtains, after renormalization, a finite effective potential~\cite{GN,Caldas2}

\begin{eqnarray}
V_{eff}(\Delta)=\frac{\Delta^2}{2 \hbar v_F } \left(\frac{1}{\lambda_{\rm GN}} - \frac{3}{2 \pi} \right) + \frac{\Delta^2}{2 \pi \hbar v_F } \ln \left( \frac{\Delta}{m_F} \right),
\label{poteff2}
\end{eqnarray}
where $\lambda_{\rm GN}$ is the renormalized coupling constant, and $m_F$ is an arbitrary renormalization scale, with dimension of energy. The minimization of $V_{eff}(\Delta)$ with respect to $\Delta$ gives the well-known result for the non-trivial gap~\cite{GN}

\begin{equation}
\Delta_0 = m_F e^{ 1- \frac {\pi}{\lambda_{\rm GN} }}.
\label{delta0}
\end{equation}
From this gap equation it is easy to see that with the experimentally measured $\Delta_0$ (and for a given $\lambda_{\rm GN}$), one sets the value of $m_F$. Equation~(\ref{poteff2}) can be expressed in a more compact form in terms of $\Delta_0$ as

\begin{eqnarray}
\label{poteff2_1}
V_{eff}(\Delta)=\frac{\Delta^2}{4 \pi \hbar v_F } \left[ \ln \left( \frac{\Delta^2}{\Delta_0^2} \right)-1 \right],
\end{eqnarray}
which is clearly symmetric under $\Delta \to -\Delta$, which generates the discrete chiral symmetry of the GN model. As has been pointed out before~\cite{Dashen}, this discrete symmetry is dynamically broken by the non-perturbative vacuum, and thus there is a kink solution interpolating between the two degenerate minima $\Delta = \pm \Delta_0$ of (\ref{poteff2_1}) at $x = \pm \infty$:

\begin{equation}
\Delta(x)=\Delta_0 \tanh(\Delta_0 x).
\label{kink}
\end{equation}

\subsubsection{Zero Temperature and Finite Chemical Potentials}

At finite chemical potentials and in the zero temperature limit, Eq.~(\ref{poteff}) reads

\begin{eqnarray}
V_{eff}(\Delta,\mu_{\uparrow,\downarrow})=\frac{1}{  \hbar v_F \lambda_{\rm GN}} \Delta^2 -   \int^{\Lambda}_0{\frac{dp}{ \pi \hbar}}~  E_p
+\int_0^{p_{F}^\uparrow} \frac{dp}{2 \pi \hbar} E_\uparrow^-  + \int_0^{p_{F}^\downarrow} \frac{dp}{2 \pi \hbar} E_\downarrow^-,
\label{poteff-4}
\end{eqnarray}
where $p_{F}^{\uparrow,\downarrow}=\frac{1}{v_{F}}\sqrt{\mu_{\uparrow,\downarrow}^2-\Delta^2}$ is the Fermi momentum of the spin-$\uparrow$($\downarrow$) moving electron. We integrate in $p$, observing that the renormalization is the same as before, to obtain

\begin{eqnarray}
V_{eff}(\Delta,\mu_{\uparrow,\downarrow})=V_{eff}(\Delta) + \Theta_1 {\cal F}_{1} + \Theta_2 {\cal F}_{2},
\label{potefff}
\end{eqnarray}
where $V_{eff}(\Delta)$ is given by Eq.~(\ref{poteff2_1}), $\Theta_{1,2}=\Theta(\mu_{\uparrow,\downarrow}^2-\Delta^2)$ is the step function, defined as $\Theta(x)=0$, for $x<0$, and $\Theta(x)=1$, for $x>0$, and ${\cal F}_{1,2} \equiv \frac{1}{2\pi \hbar v_F } \left[ \Delta^2 \ln \left(\frac{\mu_{\uparrow,\downarrow} + \sqrt{\mu_{\uparrow,\downarrow}^2-\Delta^2}}{\Delta} \right) -\mu_{\uparrow,\downarrow} \sqrt{\mu_{\uparrow,\downarrow}^2-\Delta^2} \right]$.

Extremizing $V_{eff}(\Delta,\mu_{\uparrow,\downarrow})$ with respect to $\Delta$ yields the trivial solution ($\Delta=0$), and

\begin{equation}
\label{min2}
\ln \left( \frac{\Delta}{\Delta_0} \right) +  \frac{\Theta_1}{2} {\cal G}_1 + \frac{\Theta_2}{2} {\cal G}_2  =0,
\end{equation}
where ${\cal G}_{1,2}\equiv \ln \left(\frac{\mu_{\uparrow,\downarrow} + \sqrt{\mu_{\uparrow,\downarrow}^2-\Delta^2}}{\Delta} \right)$, and we have made use of Eq.~(\ref{delta0}) to eliminate $m_F$ in the gap equation above.

The ground state is determined by jointly finding the solutions of the equation above, i.e., $\Delta_{min}(\delta \mu)$, and the analysis of the effective potential at the minimum, $V_{eff}(\Delta_{0}(\delta \mu))$. When $\delta \mu=0$ (i.e., when $B_0=0$), the solution $\Delta_{min}=\Delta_0$ represents a minimum of $V_{eff}(\Delta,\mu)$ while $\mu < \mu_c$ (where the critical chemical potential, $\mu_c=\Delta_0/\sqrt{2}$, is obtained by $V_{eff}(\Delta=0,\mu_{c})=V_{eff}(\Delta=\Delta_0,\mu_c)$), and $\Delta=0$ is a local maximum. When $\mu \geq \mu_c$, $\Delta(\mu \geq \mu_c)$ is a local maximum and $\Delta=0$ is turned into a minimum through a first-order phase transition~\cite{Wolff,CM}, agreeing with experiment~\cite{Fernando}. Thus, the gap as a function of the chemical potential in the absence of $B_0$ has the following expression:

\begin{equation}
\Delta_0(\mu)=\Theta(\mu_c-\mu) \Delta_0.
\label{gapmu}
\end{equation}

The situation changes considerably when $B_0$ is turned on, where $\Delta=0$ still represents the minimum of $V_{eff}(\Delta,\mu_{\uparrow,\downarrow})$ for $\delta \mu < \delta \mu_c = 0.372 \Delta_0$, where $\delta \mu_c$ is obtained from the equality $V_{eff}(\Delta=0,\mu_{\uparrow,\downarrow}(\delta \mu_c))=V_{eff}(\Delta=\Delta_0(\delta \mu_c),\mu_{\uparrow,\downarrow}(\delta \mu_c))$, with $\Delta_0(\delta \mu_c)$ being the solution of Eq.~(\ref{min2}) for $\mu_{\uparrow,\downarrow}(\delta \mu_c)$. For $\delta \mu \geq \delta \mu_c$, the $\Theta_{2}$ function prevents the ``$\downarrow$"-part of $V_{eff}$ in participating of the effective potential\footnote{The fact that $\Theta_{2}=\Theta(\mu_{\downarrow}^2-\Delta^2)=0$ for $\mu=\mu_c$ and $\delta \mu = \delta \mu_c$ (and beyond) can be easily verified by simply plugging in $\Theta_{2}$: $\mu_c/\Delta_0=1/\sqrt{2}$, $\delta\mu/\Delta_0=0.372$ and $\Delta=\Delta_{min} = \Delta_0^2/2\mu_\uparrow$.}. In this case, Eq.~(\ref{min2}) reduces to

\begin{equation}
\label{min4}
\Delta^4-2\mu_\uparrow \Delta_0^2 \Delta + \Delta_0^4=0.
\end{equation}

Since for $\mu=\mu_c$ the solution of the above equation is always less than $\Delta_0$ for any $\delta \mu>0$, it may be approximated as

\begin{equation}
\label{mingap}
\Delta_{min} \simeq \frac{\Delta_0^2}{2\mu_\uparrow}.
\end{equation}

Notice that for $\mu < \mu_c$ and $\delta \mu < \delta \mu_c$ the dispersion of the fermions is $E_{\uparrow,\downarrow}^{\pm} = \sqrt{v_F^2 p^2+\Delta_0^2} \pm \mu_{\uparrow,\downarrow}$, which is the typical energy of massive fermions under a Zeeman magnetic field;
\newline 
For $\mu \geq \mu_c$ and $\delta \mu < \delta \mu_c$ the dispersion of the fermions is linear i.e., $E_{\uparrow,\downarrow}^{\pm} = v_F p \pm \mu_{\uparrow,\downarrow}$, which is a signature of gapless (massless) systems~\cite{CastroNeto}, and finally;
\newline
For $\mu \geq \mu_c$ and $\delta \mu \geq \delta \mu_c$ the fermions energies are given by $E_{\uparrow,\downarrow}^{\pm} = \sqrt{v_F^2 p^2+\Delta_{min}^2} \pm \mu_{\uparrow,\downarrow}$, which is again the dispersion of massive fermions subjected to an external Zeeman magnetic field.

\begin{figure}[h]
\centering
\includegraphics[height=4.0in]{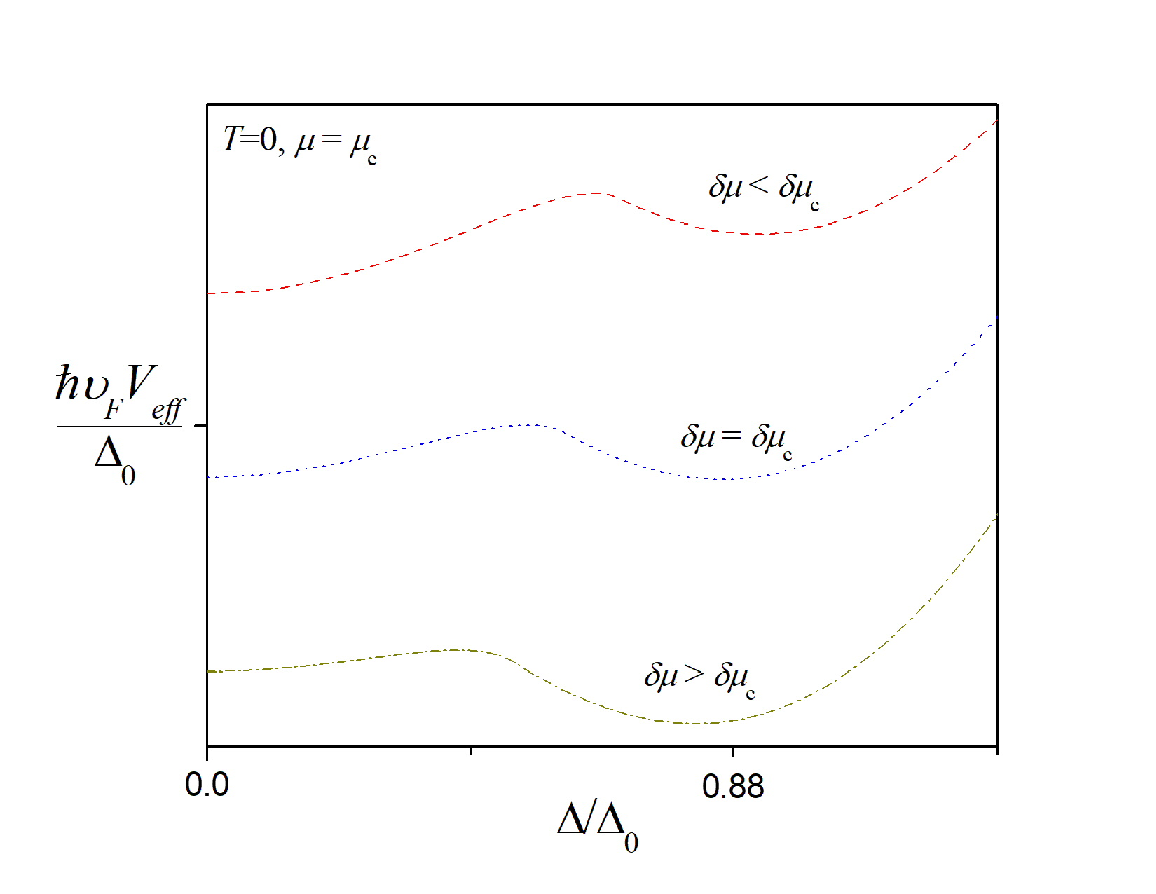}
\caption{\label{gapped}The zero temperature effective potential $V_{eff}$ from Eq.~(\ref{potefff}) for various strengths of Zeeman magnetic field. All curves are for $\mu = \mu_c$ and $\delta \mu$ varies, from top to bottom, as $\delta \mu < \delta \mu_c$, $\delta \mu = \delta \mu_c$, and $\delta \mu > \delta \mu_c$.}
\end{figure}

Let us now describe the emergence of this fully polarized state at zero temperature. Suppose the system was in a gapless phase with $\mu=\mu_c$ and $\delta \mu=0$. The magnetic field is then applied parallel to the system and increased. As $\delta \mu$ reaches the critical value $\delta \mu_c=0.372 \Delta_0$, or 
\begin{equation}
B_{0,c}= 0.744 \frac{\Delta_0}{g \mu_B},
\label{cmf2}
\end{equation}
there is a quantum phase transition to a gapped phase, with the gap $\Delta_{min}$, which is the minimum of $V_{eff}$, given as a solution of Eq.~(\ref{min4}), i.e., by Eq.~(\ref{mingap})~\cite{Caldas2}. This behavior can be seen in Fig.~\ref{gapped}, which clearly shows that the effect of a Zeeman field $\delta \mu$ is to induce a phase transition from a gapless to a gapped phase at $\delta \mu_c$.

This suppression of the gapless phase is analogous to the one found in a 2D system, where the Zeeman magnetic field suppresses the metallic behavior, inducing a metal-insulator transition at a critical field strength~\cite{Dent}.

\begin{figure}[h]
\centering
\includegraphics[height=4.0in]{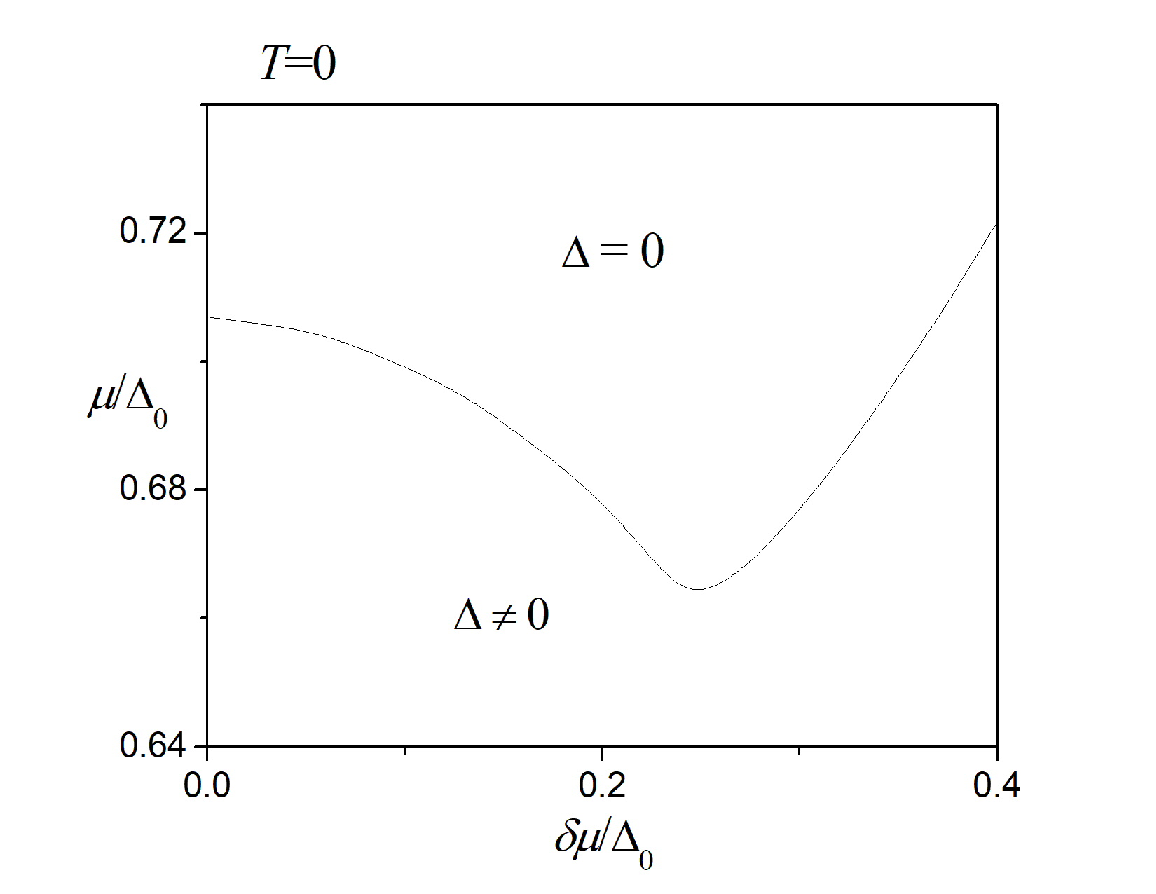}
\caption{\label{zeroTpd} Zero temperature phase diagram in the $\mu/\Delta_0$ versus $\delta \mu/\Delta_0$ plane. Notice, for instance, that for $\mu/\Delta_0 = \mu_c/\Delta_0 = 1/\sqrt{2}\approx 0.707$ (which restores the symmetry at $\delta\mu/\Delta_0=0$), and after increasing the magnetic field, there is a first-order phase transition from the gapless to the gapped phase for $\delta\mu/\Delta_0 = 0.372$, see Fig.~\ref{gapped}.}
\end{figure}

In Fig.~\ref{zeroTpd}, we present a zero temperature phase diagram showing gapped and gapless phases as a function of both $\mu$ and $\delta \mu$. For $\delta \mu=0$, starting with $\mu=0$, we are in the symmetry-broken phase, and as we increase $\mu$, we reach the critical chemical potential $\mu_c/\Delta_0=1/\sqrt{2}$ at which there is the well-known first-order phase transition to the gapless phase. However, there are interesting new phase transitions due to the actuation of the Zeeman magnetic field in the system. If we fix, for instance, $\mu=\mu_c$ and turn on the Zeeman field, increasing $\delta \mu$ there is a first-order phase transition to the gapped phase at $\delta\mu_c/\Delta_0=0.372$ that we just discussed above. Starting now with some $\mu < \mu_c$, and turning on the Zeeman field, as we increase $\delta \mu$ there is a ``reentrant'' behavior with gapped-gapless-gapped phases.

It is important to notice that the critical field $\delta \mu_c$ (or $B_{0,c}$ in Eq.~(\ref{cmf2})), which drives a transition to a polarized electronic system, can be dramatically decreased as a consequence of the enhancement of the $g$-factor due to carrier doping. See Ref.~\cite{Xuan}, where an ab initio approach, based on many-body perturbation theory, to compute the interaction-enhanced Land\'e $g$-factors in carrier-doped systems has been developed.

The number densities $n_{\uparrow,\downarrow}= -\frac{\partial}{\partial \mu_{\uparrow,\downarrow}} V_{eff}(\Delta,\mu_{\uparrow,\downarrow})$ are obviously imbalanced due to the asymmetry between $\mu_\uparrow$ and $\mu_\downarrow$, and will depend on $\delta \mu$. Before the critical magnetic field, $\Delta_0(\delta \mu)=0$ and the densities read

\begin{equation}
n_{\uparrow,\downarrow}(\delta \mu/\Delta_0 < 0.372)= \frac{1}{\pi \hbar v_F } \mu_{\uparrow,\downarrow}.
\label{nd}
\end{equation}
For such a low imbalance, compared to the critical chemical potential asymmetry $\delta \mu_c$, the total number density is the same as in the symmetric limit, or, in other words, is independent of the applied magnetic field,

\begin{equation}
n_T=n_{\uparrow}+n_{\downarrow}=2n=\frac{2}{\pi \hbar v_F } \mu_c. 
\label{Nd1}
\end{equation}
However, we can obtain a partial spin polarization

\begin{equation}
\delta n=n_{\uparrow}- n_{\downarrow}=\frac{2}{\pi \hbar v_F } \delta \mu = \frac{2}{\pi \hbar v_F } \mu_B B_0,
\label{Nd2}
\end{equation}
where we have considered for the moment $g \approx 2$ such that $\delta \mu =|\Delta E|=\mu_B B_0$.

Increasing the asymmetry beyond the critical value, we find

\begin{equation}
n_{\uparrow}(\delta \mu/\Delta_0 > 0.372)= \frac{1}{\pi \hbar v_F }\sqrt{\mu_{\uparrow}^2-\Delta^2},
\label{Nd3}
\end{equation}
where $\Delta$ in the equation above is the solution of Eq.~(\ref{min4}) for a given $\delta \mu$, and

\begin{equation}
n_{\downarrow}(\delta \mu/\Delta_0 > 0.372)= 0,
\label{Nd4}
\end{equation}
meaning that, effectively, there are only spin-$\uparrow$ electrons in the system i.e., it is fully polarized. The number difference is now

\begin{equation}
\delta n=n_{\uparrow},
\label{nd5}
\end{equation}
which also depends on the external magnetic field $B_0$ through $\mu_{\uparrow}$ and $\Delta$. As we have just mentioned, this polarization of the electronic system, i.e., having electrons with only one type of spin, results in a noninteracting (insulating) system~\cite{Dent}.

\subsubsection{Finite Temperature and Chemical Potentials}

We can rewrite $V_{eff}(\Delta,\mu_{\uparrow,\downarrow},T)$ as

\begin{equation}
V_{eff}(\Delta,\mu_{\uparrow,\downarrow},T)= V_{eff}(\Delta) + V_{eff}(\mu_{\uparrow,\downarrow},T),
\label{poteff3}
\end{equation}
where $V_{eff}(\Delta)$ is given by Eq.~(\ref{poteff2_1}), and

\begin{eqnarray}
V_{eff}(\mu_{\uparrow,\downarrow},T)=
- k_B T \int^{\infty}_{0} \frac{dp}{2 \pi \hbar } ~ \Big[  \ln \left(1+e^{-\beta E_\uparrow^+}\right)+ \ln \left(1+e^{-\beta E_\uparrow^-}\right)
+ \ln \left(1+e^{-\beta E_\downarrow^+}\right) + \ln \left(1+e^{-\beta E_\downarrow^-}\right) \Big].
\label{poteff4}
\end{eqnarray}

In order to evaluate the phase structure of the system, Eq. (\ref{poteff4}) can be numerically evaluated at each temperature $T$, chemical potential $\mu$, and chemical potential asymmetry $\delta\mu$. Alternatively, to obtain, analytically, approximate expressions for the phase transition curves and tricritical points, at least in the high-temperature regime, we can use a high-temperature expansion to evaluate it~\cite{Rudnei}. Using the function

\begin{equation}
I(a,b)=\int_0^{\infty} dx \left[ \ln \left(1+ e^{-\sqrt{x^2+a^2}-b} \right) + \ln \left(1+ e^{-\sqrt{x^2+a^2}+ b} \right) \right],
\label{int1}
\end{equation}
where $a=\Delta/k_BT$, and $b=\mu/k_BT$, which can be expanded in the high temperature limit, $a<<1$ and $b<<1$, yielding, up to order $a^6$ and $b^6$,

\begin{eqnarray}
I(a<<1,b<<1)&=&\frac{\pi^2}{6}+\frac{b^2}{2}-\frac{a^2}{2}\ln \left(\frac{\pi}{a} \right) -\frac{a^2}{4}(1-2\gamma_E)-\frac{7 \zeta(3)}{2^3 \pi^2} a^2 b^2 + \frac{31 \zeta(5)}{2^5 \pi^4} a^2 b^4 - \frac{127 \zeta(7)}{2^7 \pi^6} a^2 b^6 \nonumber \\
&&-\frac{7 \zeta(3)}{2^5 \pi^2} a^4 + \frac{186 \zeta(5)}{2^7 \pi^4} a^4 b^2 - \frac{1905 \zeta(7)}{2^9 \pi^6} a^4 b^4 -\frac{14308 \zeta(9)}{2^{11} \pi^8} a^4 b^6 \nonumber \\
&&+\frac{62 \zeta(5)}{2^9 \pi^4} a^6 - \frac{3810\zeta(7)}{2^{11} \pi^6} a^6 b^2 + \frac{71540\zeta(9)}{2^{13} \pi^8} a^6 b^4 - \frac{859740 \zeta(11)}{2^{15} \pi^{10}} a^6 b^6+ {\cal O}\left(a^8, b^8 \right),
\end{eqnarray}
where $\gamma_E \approx 0.577... $ is the Euler constant and $\zeta(n)$ is the Riemann zeta function, having the values $\zeta(3) \approx 1.2021$, $\zeta(5) \approx 1.0369$, $\zeta(7) \approx 1.0083$, $\zeta(9) \approx 1.0020$ and $\zeta(11) \approx 1.0005$. With the equation above, together with Eq.~(\ref{poteff2_1}), the high temperature ``asymmetrical'' effective potential can be rearranged in the form of a Ginzburg-Landau (GL) expansion of the grand potential density, which is appropriate to the analysis of the phase diagram in the high-temperature region, given by

\begin{equation}
\label{Veff}
V_{eff}(\mu_{\uparrow,\downarrow},T)= \alpha_0 + \alpha_2 \Delta^2 + \alpha_4 \Delta^4 + \alpha_6 \Delta^6,
\end{equation}
where
\begin{equation}
\label{coef}
\alpha_0(\mu_{\uparrow,\downarrow},T) = - d \left(\frac{ 2 \pi^2}{6} + \frac{1}{2} \frac{(\mu_{\uparrow}^2+\mu_{\downarrow}^2)}{(k_B T)^2}   \right) (k_B T)^2, 
\end{equation}

\begin{equation}
\alpha_2(\mu_{\uparrow,\downarrow},T) = -d \left( -\ln \left( \frac{\pi k_B T}{e^{\gamma_E} \Delta_0} \right) - \frac{7 \zeta(3)}{2^3 \pi^2} \frac{(\mu_{\uparrow}^2+\mu_{\downarrow}^2)}{(k_B T)^2}
+ \frac{31 \zeta(5)}{2^5 \pi^4} \frac{(\mu_{\uparrow}^4+\mu_{\downarrow}^4)}{(k_B T)^4}
- \frac{127 \zeta(7)}{2^7 \pi^6} \frac{(\mu_{\uparrow}^6+\mu_{\downarrow}^6)}{(k_B T)^6} \right), 
\end{equation}
\begin{equation}
\alpha_4(\mu_{\uparrow,\downarrow},T) = - d \left( - \frac{14 \zeta(3)}{2^5 \pi^2} + \frac{186 \zeta(5)}{2^7 \pi^4} \frac{(\mu_{\uparrow}^2+\mu_{\downarrow}^2)}{(k_B T)^2} 
- \frac{1905 \zeta(7)}{2^9 \pi^6} \frac{(\mu_{\uparrow}^4+\mu_{\downarrow}^4)}{(k_B T)^4} 
- \frac{14308 \zeta(9)}{2^{11} \pi^8} \frac{(\mu_{\uparrow}^6+\mu_{\downarrow}^6)}{(k_B T)^4} \right) \frac{1}{(k_B T)^2},
\end{equation}
and
\begin{equation}
\label{alpha6}
\alpha_6(\mu_{\uparrow,\downarrow},T) = -d \left( \frac{124 \zeta(5)}{2^9 \pi^4} - \frac{3810 \zeta(7)}{2^{11} \pi^6} \frac{(\mu_{\uparrow}^2+\mu_{\downarrow}^2)}{(k_B T)^2} 
+ \frac{71540 \zeta(9)}{2^{13} \pi^8} \frac{(\mu_{\uparrow}^4+\mu_{\downarrow}^4)}{(k_B T)^4} 
- \frac{859740 \zeta(11)}{2^{15} \pi^{10}} \frac{(\mu_{\uparrow}^6+\mu_{\downarrow}^6)}{(k_B T)^6} 
\right) \frac{1}{(k_B T)^4},
\end{equation}
where $d \equiv \frac{1}{2 \pi \hbar v_F}$.

It is interesting to notice that Eq.(\ref{Veff}) taken up to the order $\mu^2_{\uparrow,\downarrow}/T^2$ reproduces qualitatively the same phase structure of the numerical evaluation of Eq. (\ref{poteff4}). The analysis of this approximate effective potential allows one to determine second- and first-order transition curves (at least for $T \geq 0.5 T_c(0)$) and the tricritical point, characterized by a tricritical temperature $T_{tc}$ and tricritical chemical potential $\mu_{tc}$, up to this order of approximation. In what follows, we will study the phase transition structure in this approximation, and we will discuss errors and higher-order terms later.

\begin{figure}[h]
\centering
\includegraphics[height=4.0in]{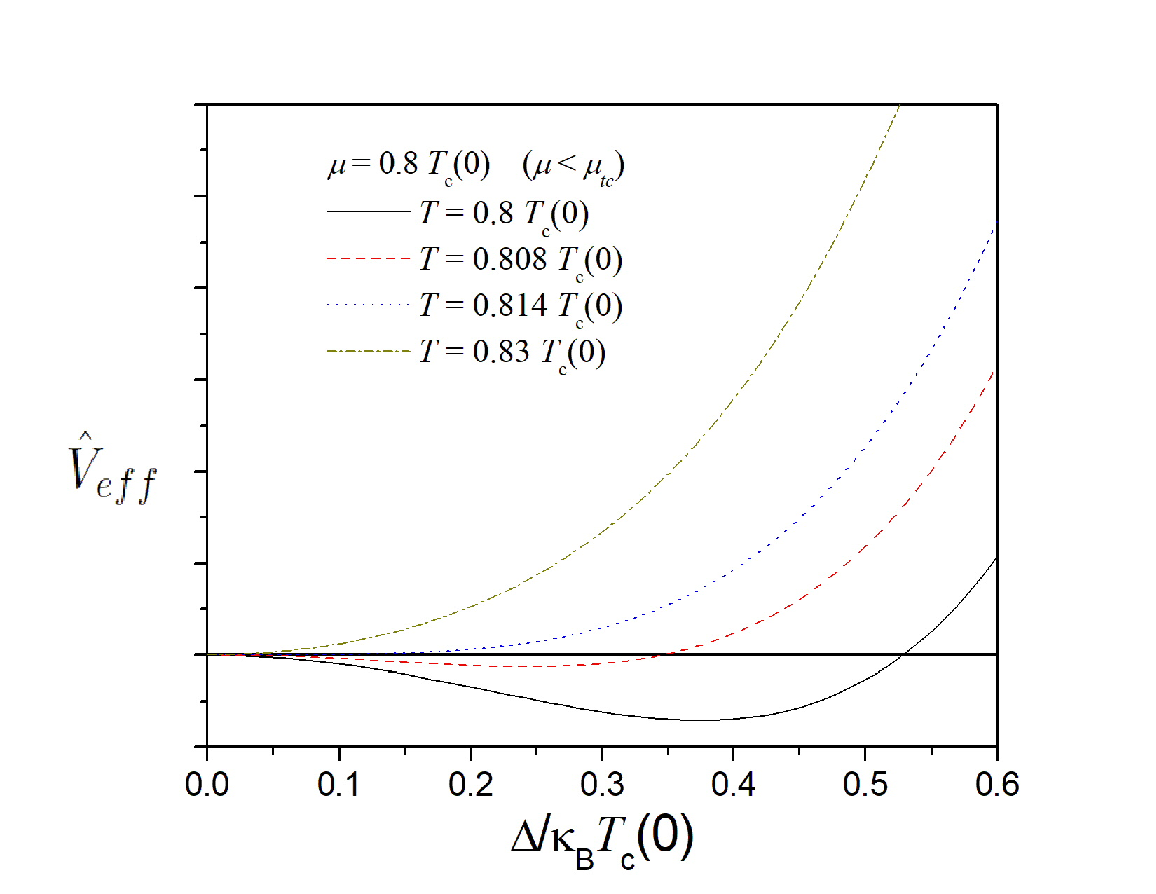}
\caption{\label{Pot-1} 
(color online) The effective potential as a function of the gap ($\Delta$) for fixed $\mu < \mu_{tc}$ at different temperatures, with $\delta \mu = 0$. For instance, for $\mu = 0.8 T_{C}(0)$ and $T=0.8T_{C}(0)$ (solid black line), the system presents one single stable gap at $\Delta/T_{C}(0) \approx 0.38$. As the temperature increases, the stable minimum shifts to lower values, e.g., $\Delta/T_{C}(0) \approx 0.245$ (dashed red line), for $T = 0.808 T_{C}(0)$. For $T \ge 0.814 T_{C}(0)$, the system presents only one stable gap at $\Delta=0$ (dotted blue and dash-dotted olive lines).}
\end{figure}

\begin{figure}[h]
\centering
\includegraphics[height=4.0in]{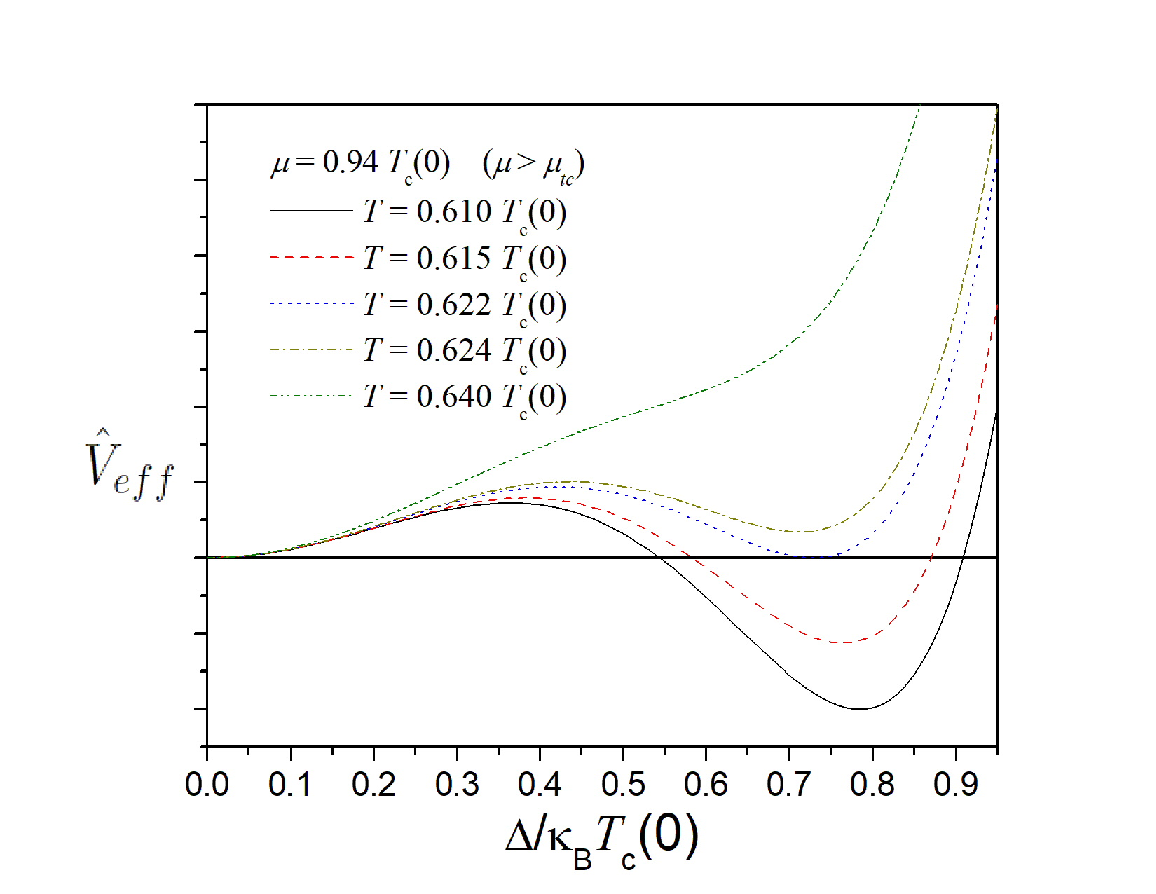}
\caption{\label{Pot-2}
(color online) The effective potential as a function of the gap ($\Delta$) for fixed $\mu > \mu_{tc}$ at different temperatures, with $\delta \mu = 0$. For instance, at $\mu = 0.94 T_{c}(0)$ and $T=0.610 T_{c}(0)$ (solid black line), the effective potential presents two minima: one (metastable) minimum located at $\Delta=0$ and the second, more pronounced (stable), located at $\Delta \ne 0$. As $T$ increases, the effective potential at the second minimum approaches $V_{eff}(\Delta=0)$ (dashed red line). At $T \approx 0.622 T_{c}(0)$, the effective potential has the same value at the two minima, characterizing the first-order transition temperature at $\mu = 0.94 T_{c}(0)$ and $\delta\mu=0$. For higher temperatures (e.g., $T = 0.624 T{c}(0)$, dash-dotted olive line), the minimum at $\Delta=0$ is stable, and the second minimum becomes meta-stable. For even higher temperatures ($T = 0.64 T_{c}(0)$, for instance, dash-dot-dotted line), the effective potential exhibits only one single stable minimum at $\Delta = 0$.}
\end{figure}

In figures~\ref{Pot-1} and~\ref{Pot-2} we show the non-dimensional effective potential $\hat{V}_{eff} \equiv \hbar v_F\frac{V_{eff}(\Delta)-V_{eff}(0)}{\Delta_0}$, up to order $\mu^2_{\uparrow,\downarrow}/T^2$, for various temperatures at fixed chemical potentials and zero external magnetic field. Here, $T_c(0)$ is the critical temperature at zero external magnetic field, to be derived later (Eq.(\ref{Tc2})). Fig.~\ref{Pot-1} is for $\mu < \mu_{tc}$, while Fig.~\ref{Pot-2} is for $\mu > \mu_{tc}$.

Minimizing $V_{eff}$ we find the trivial solution ($\Delta=0$) and the chemical potential and temperature dependent gap equation
\begin{equation}
\label{Deltamin}
\Delta(\mu_{\uparrow,\downarrow},T)^2= -\frac{\alpha_4}{3 \alpha_6} \left(1 \pm \sqrt{1-\frac{3 \alpha_2 \alpha_6}{\alpha_4^2}} \right),
\end{equation}
where the relative sign determines whether the extreme is a maximum or a minimum, depending on the sign of $\alpha_4$ and $\alpha_6$. For high temperatures we can neglect the $\alpha_6$ term to study the second order phase transition, obtaining

\begin{equation}
\label{poteff6}
\Delta(\mu_{\uparrow,\downarrow},T)^2= -\frac{\alpha_2}{2 \alpha_4},
\end{equation}
which has meaning only if the ratio $\frac{\alpha_2}{ \alpha_4}$ is negative. Besides, a stable configuration (i.e., bounded from below) requires, up to this order, $\alpha_4>0$. At the minimum $V_{eff}$ reads

\begin{equation}
\label{Veff2}
V_{eff, min}= \alpha_0 - \frac{\alpha_2^2}{4 \alpha_4}.
\end{equation}

The critical temperature $T_c$ is, by definition, the temperature at which the gap vanishes. Thus, at $T_c$ we have $\alpha_2=0$, or, up to order $(\mu^2{\uparrow}+\mu^2{\downarrow})/T^2$,

\begin{equation}
 \ln \left( \frac{\pi k_B T_c}{e^{\gamma_E} \Delta_0} \right) + \frac{7 \zeta(3)}{8 \pi^2} \frac{(\mu_{\uparrow}^2+\mu_{\downarrow}^2)}{(k_B T_c)^2} =0,
\label{Tc1}
\end{equation}
where $T_c=T_c(\mu_{\uparrow},\mu_{\downarrow})$. As will become clear below, the equation above defines a second order transition line separating the gapped ($\Delta \neq 0$) and gapless phases ($\Delta = 0$). At $\mu_{\uparrow}=\mu_{\downarrow}=0$, we recover the well-known result for the temperature at which the discrete chiral symmetry is restored~\cite{Jacobs}

\begin{equation}
T_c(\mu_{\uparrow}=\mu_{\downarrow}=0) \equiv T_c(0) = \frac{\Delta_0}{k_B} \frac{e^{\gamma_E}}{\pi}.
\label{Tc2}
\end{equation}

In order to find an analytical expression for the critical temperature as a function of both $\mu_{\uparrow}$ and $\mu_{\downarrow}$ i.e., $T_c(\mu_{\uparrow},\mu_{\downarrow})$, from Eq.~(\ref{Tc1}) we write

\begin{equation}
T_c(\mu_{\uparrow},\mu_{\downarrow}) = \frac{\Delta_0}{k_B} \frac{e^{\gamma_E}}{\pi} e^{-\frac{7 \zeta(3)}{8 \pi^2} \frac{(\mu_{\uparrow}^2+\mu_{\downarrow}^2)}{(k_B T_c)^2}} .
\label{Tcmu1}
\end{equation}
At small $\mu_{\uparrow}^2+\mu_{\downarrow}^2$, we may approximate

\begin{equation}
T_c(\mu_{\uparrow},\mu_{\downarrow}) = \frac{\Delta_0}{k_B} \frac{e^{\gamma_E}}{\pi} \left[1-\frac{7 \zeta(3)}{8 } \frac{(\mu_{\uparrow}^2+\mu_{\downarrow}^2)}{\Delta_0^2 e^{2\gamma_E}} \right],
\label{Tcmu2}
\end{equation}
which is the phase transition temperature in the regime where the transition is of second order. Thus, this transition temperature is valid along the second-order curve until the tricritical point, which we will find below. Notice that in the limit $\delta \mu=0$ in Eq.~(\ref{Tcmu2}), we obtain the result in Ref.~\cite{Chodos}.

As we can see from Fig. \ref{Pot-2}, for low enough temperatures and high enough chemical potentials, the effective potential presents two minima, one located at $\Delta=0$ and the other at some finite value ($\Delta_{min}$), given by Eq. (\ref{Deltamin}). In this regime, the $\alpha_6$ term cannot be neglected anymore. For fixed chemical potential, as we increase the temperature, the stable (lower) minimum switches from $\Delta = \Delta_{min}$ to $\Delta = 0$. The critical temperature is characterized, thus, by $V_{eff}(\Delta=0) = V_{eff}(\Delta_{min}(\mu_{\uparrow,\downarrow},T_c))$. By replacing Eq. (\ref{Veff}) in this last equation, we obtain
\begin{equation}
\alpha_2 + \alpha_4 \Delta^2_{min}(\mu_{\uparrow,\downarrow},T_c) + \alpha_6 \Delta^4_{min}(\mu_{\uparrow,\downarrow},T_c) = 0,
\end{equation}

or
\begin{equation}
\Delta_{min}^2(\mu_{\uparrow,\downarrow},T_c)= -\frac{\alpha_4}{2 \alpha_6} \left(1 \pm \sqrt{1-\frac{4 \alpha_2 \alpha_6}{\alpha_4^2}} \right). \label{Deltamin1storder}
\end{equation}
Eqs.(\ref{Deltamin}) and (\ref{Deltamin1storder}) can only be simultaneously satisfied if 
\begin{equation}
\alpha_4^2 - 4 \alpha_2 \alpha_6 = 0,
\end{equation}
the equation that characterizes the first-order phase transition curves, to be solved for $T_c$ as a function of the chemical potentials $\mu_{\uparrow}$ and $\mu_{\downarrow}$.

\begin{figure}[h]
\centering
\includegraphics[height=4.0in]{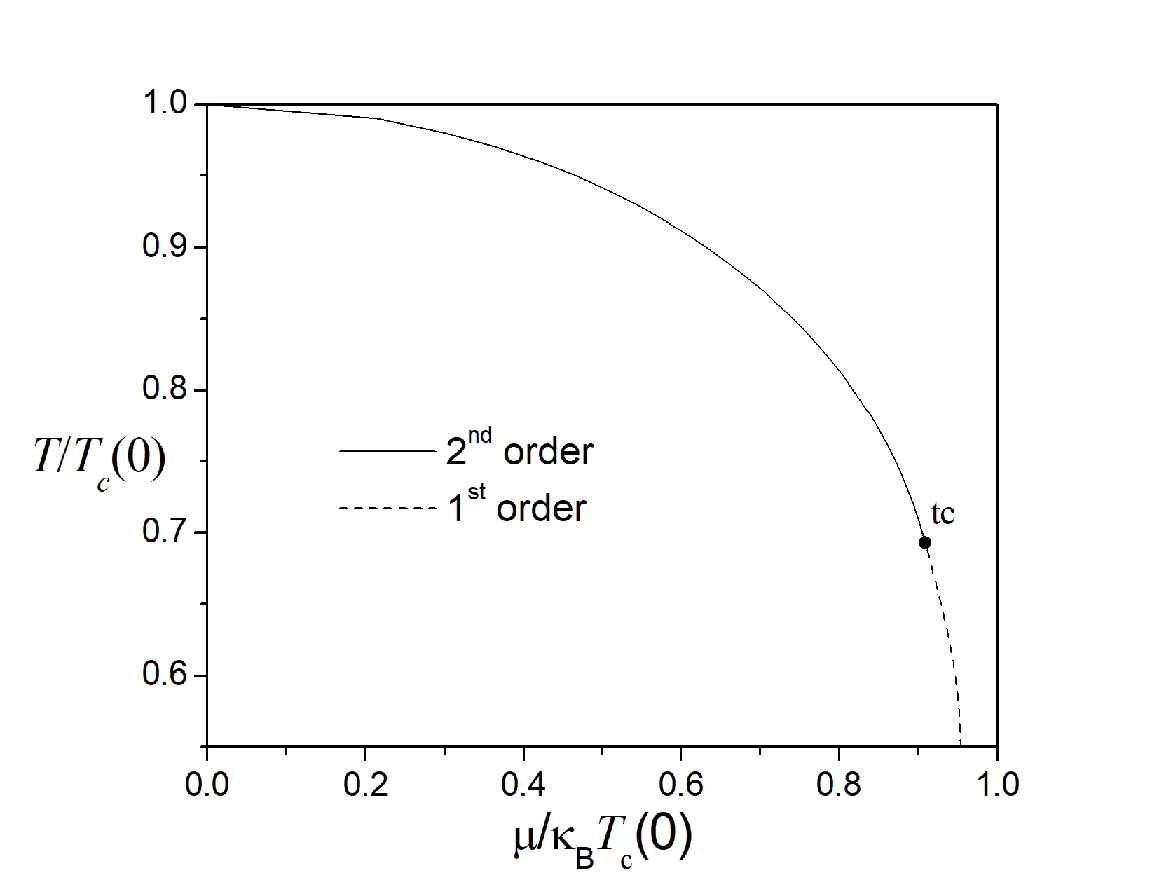}
\caption{\label{PDzero} The phase diagram for the Gross-Neveu model expanded up to the second order in the chemical potentials. The solid line is the continuous second-order phase transition curve to the gapless phase at high temperatures or chemical potentials. The dashed line indicates the first-order phase transition line, where the gap abruptly jumps from $\Delta \ne 0$ to $\Delta=0$. The tricritical point (tc), at $t=0.69274$ and $\mu/T_c(0)=0.90914$, is indicated in the figure as the black dot.}
\end{figure}

\begin{figure}[h]
\centering
\includegraphics[height=4.0in]{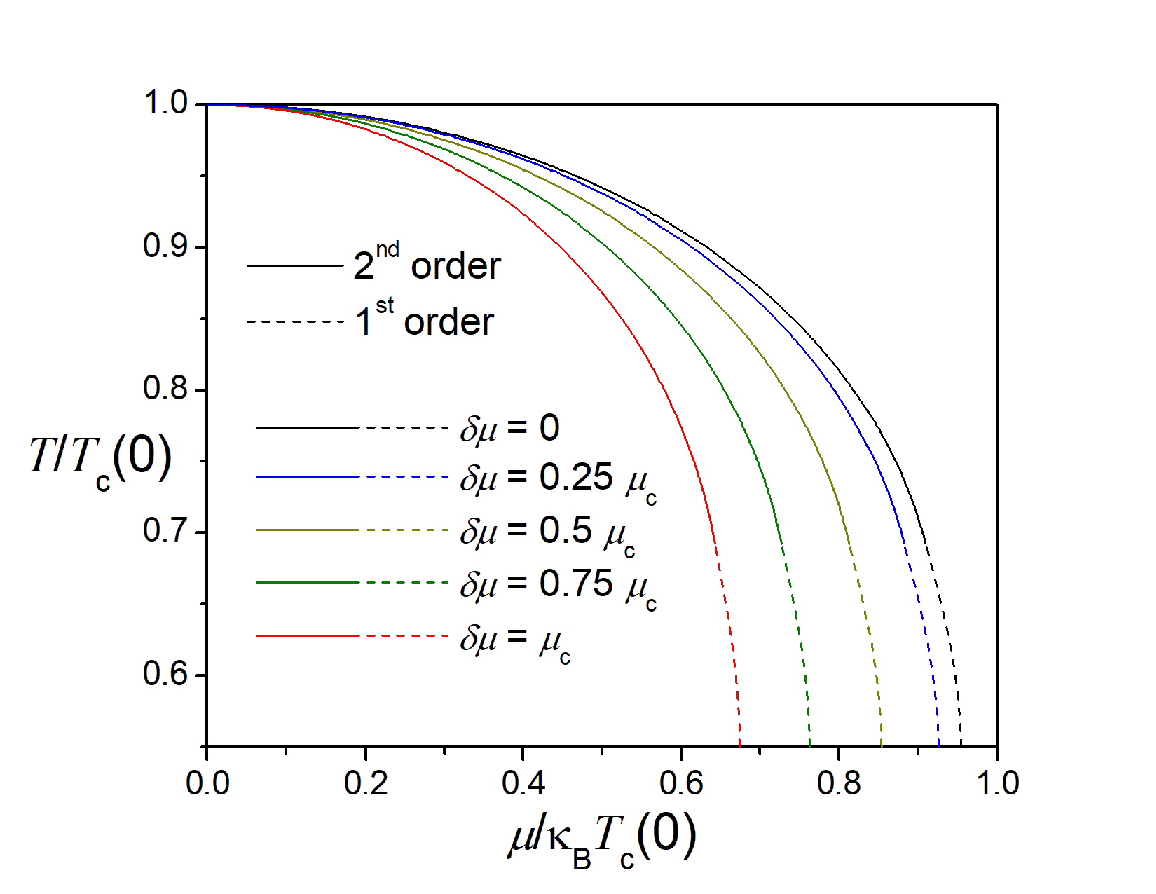}
\caption{\label{distinct} (color online) Phase diagrams for distinct chemical potential asymmetries, up to the second-order in $\mu_{\uparrow,\downarrow}/T$. Solid and dashed lines are the second-order and first-order phase transition curves, respectively. The tricritical chemical potential decreases (while the tricritical temperature does not vary) as the chemical potential asymmetry grows. For $\delta \mu = \mu_c$, the system is in the fully polarized state ($\mu_{\downarrow} = 0$).}
\end{figure}

Finally, the tricritical point is the point where the second-order transition curve reaches the first-order one, i.e., at the tricritical point both conditions for the second-order transition ($\alpha_2=0$) and first-order transition ($\alpha_4^2 - 4 \alpha_2 \alpha_6 = 0$) have to be fulfilled. At the tricritical point, these two equations are satisfied if $\alpha_2=0$ and $\alpha_4=0$.
Numerically solving this set of two equations (up to the second-order in $\mu_{\uparrow,\downarrow}/T$ and at zero external magnetic field), we find $T_{tc} = 0.6927 T_{c}(0) = 0.3927 \Delta_0$ and $\mu_{tc} = 0.9091 T_{c}(0) = 0.5154 \Delta_0$, the tricritical point temperature and chemical potential, respectively, within the high temperature expansion we employed. This result should be compared to the numerical results \cite{Wolff}, $T_{tc} = 0.318 \Delta_0$ and $\mu_{tc} = 0.608 \Delta_0$. The quantitative difference is due to the high-temperature expansion we employed. Qualitative results, however, coincide with previous ones obtained at zero external magnetic field \cite{Wolff,Lenz2} and are shown in Fig. \ref{PDzero}. Figure \ref{distinct} shows the phase diagrams computed at second-order in $\mu_{\uparrow,\downarrow}/T$ with different chemical potential asymmetries. 
At this order of approximation, the tricritical temperature $T_{tc}$ does not depend on $\delta\mu$ while the tricritical chemical potential $\mu_{tc}$ does depend. Notice that as $\delta\mu$ is increased $T_{tc}$ does not vary while $\mu_{tc}$ decreases \footnote{We will see below that when higher order of the expansion on $\mu_{\uparrow,\downarrow}/T$ is taken into account $T_{tc}$ will slightly vary as $\delta\mu$ increases (see Fig.~\ref{mutcvsdmu})}. As an overall result, the gapped phase region in the phase diagram is reduced as the external magnetic field increases. 

While the second-order phase transition curve is relatively stable with respect to the high-temperature expansion, Eqs. (\ref{Veff} -- \ref{alpha6}), the same does not occur for the first-order phase transition. The reasons for that are simple: the second-order transition curve is computed at higher temperatures and in the limit $\Delta \rightarrow 0$, whereas the first-order curve, in contrast, lies in a not-so-high temperature region and depends on the effective potential computed at $\Delta/T \neq 0$, so errors of order $(\Delta/T)^6$ are also involved in this second case. As a consequence, although it is possible to find a first-order transition curve and the tricritical point when we consider high-order terms in the aforementioned expansion (up to order 6, for example), results do not converge to the exact numerically computed ones.

\begin{figure}[h]
\centering
\includegraphics[height=4.0in]{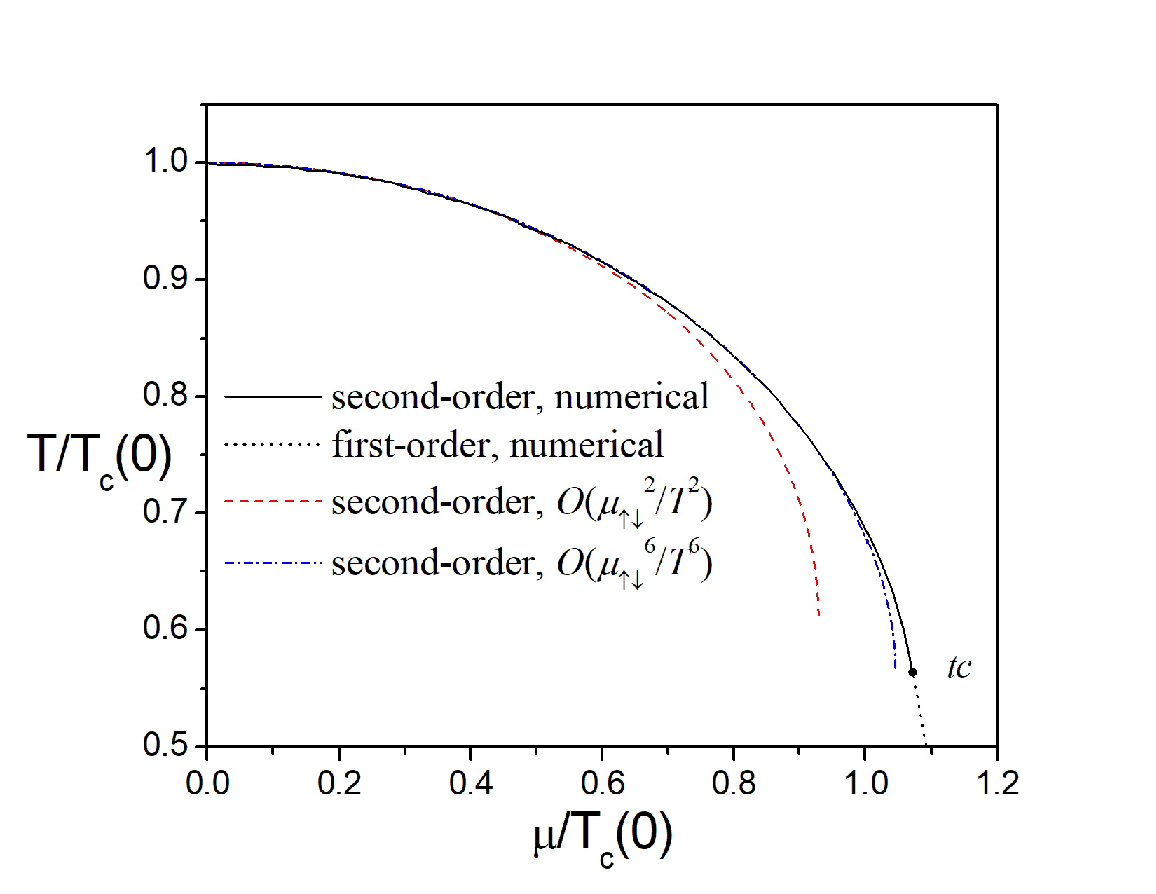}
\caption{\label{Exacto2o6} (color online) The second-order phase transition curves, computed up to order $\mu^2_{\uparrow,\downarrow}/T^2$ (dashed red line) and $\mu^6_{\uparrow,\downarrow}/T^6$ (dash-dotted blue line) in comparison with the numerically evaluated second-order phase transition curve (solid black line). The tricritical point (full black dot) and part of the first-order transition curve (dotted black line) are also displayed. The rightmost point in each high-temperature approximation for the second-order transition curves approximates the exact (numerically computed) tricritical point.}
\end{figure}

As mentioned before, errors are lower in the evaluation of the second-order transition curve. Figure \ref{Exacto2o6} shows the second-order transition curves computed up to orders $\mu^2_{\uparrow,\downarrow}/T^2$ and $\mu^6_{\uparrow,\downarrow}/T^6$ compared to the numerically computed one, in which we also locate the tricritical point. As we can see, the second-order transition curve computed up to order $\mu^6_{\uparrow,\downarrow}/T^6$ approaches the more accurate numerical result.

To approximate the tricritical point by means of analytical expressions, one can observe that the highest chemical potential for which the second-order condition $\alpha_2=0$ can be fulfilled occurs very close to the tricritical point -- in fact, at approximately the same chemical potential and a slightly lower temperature, allowing us to find an analytical approximation for the tricritical point. Following \cite{jstat}, let us define

\begin{eqnarray}
\label{y-1}
y(t) &\equiv& \alpha_2(\mu_{\uparrow,\downarrow},T) = d \left( \ln \left( \frac{\pi k_B T}{e^{\gamma_E} \Delta_0} \right) + \frac{7 \zeta(3)}{2^3 \pi^2} \frac{(\mu_{\uparrow}^2+\mu_{\downarrow}^2)}{(k_B T)^2}
- \frac{31 \zeta(5)}{2^5 \pi^4} \frac{(\mu_{\uparrow}^4+\mu_{\downarrow}^4)}{(k_B T)^4} + \frac{127 \zeta(7)}{2^7 \pi^6} \frac{(\mu_{\uparrow}^6+\mu_{\downarrow}^6)}{(k_B T)^6}
 \right),
\end{eqnarray}
where $d$ was defined below Eq.~(\ref{alpha6}).

The function $y(t)$ may be written as
\begin{eqnarray}
\label{y-2}
y(t) &=&  d \left[ \ln \left( t \right) + \frac{a}{t^2} -  \frac{b}{t^4} +  \frac{c}{t^6} \right],
\end{eqnarray}
where $t=\frac{\pi k_B T}{e^{\gamma_E} \Delta_0}=\frac{T}{T_c(0)}$,  $a=\frac{7 \zeta(3)}{2^3 \pi^2} \frac{(\mu_{\uparrow}^2+\mu_{\downarrow}^2)}{(k_B T_c(0))^2}$, $b=\frac{31 \zeta(5)}{2^5 \pi^4} \frac{(\mu_{\uparrow}^4+\mu_{\downarrow}^4)}{(k_B T_c(0))^4}$, and $c = \frac{127 \zeta(7)}{2^7 \pi^6} \frac{(\mu_{\uparrow}^6+\mu_{\downarrow}^6)}{(k_B T_c(0))^6}$. 

\begin{figure}[h]
\centering
\includegraphics[height=4.0in]{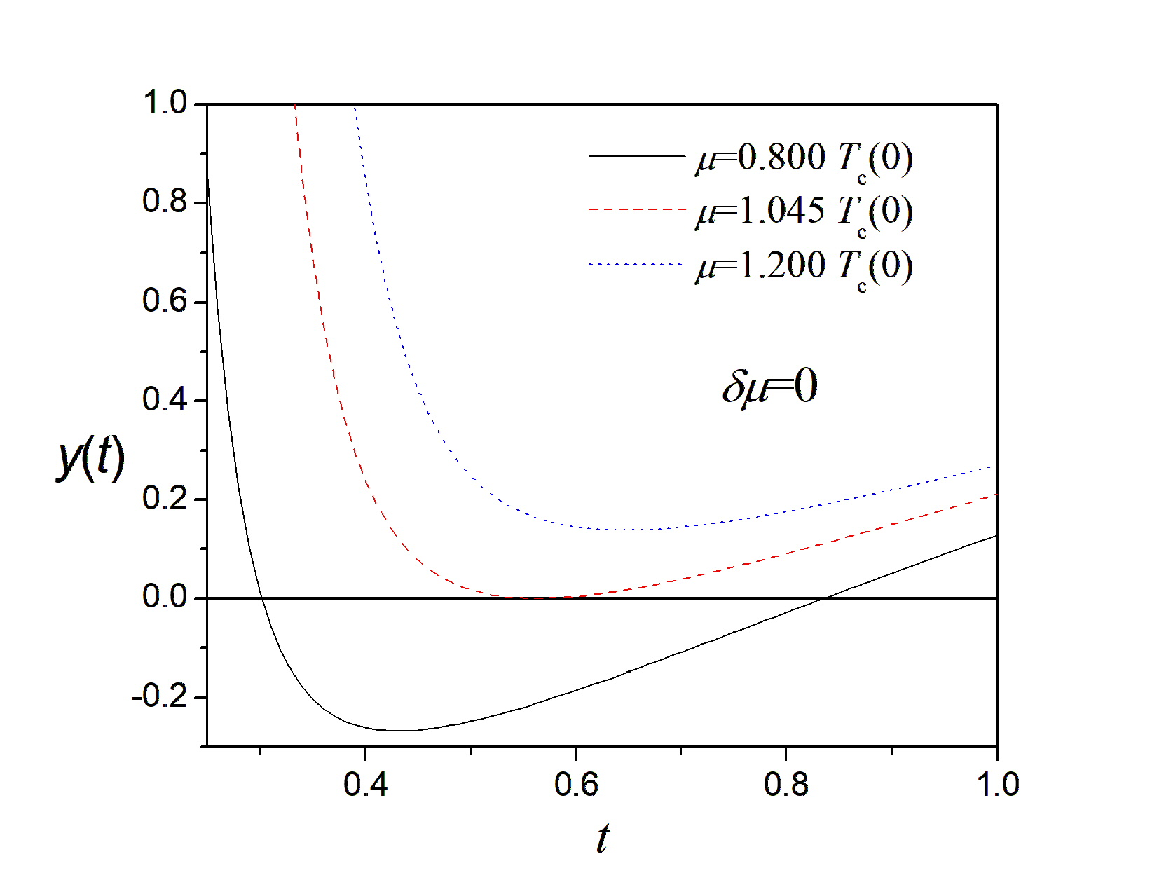}
\caption{\label{y} (color online) The function $y(t)$, as a function of $t=\frac{T}{T_c(0)}$, for different values of the chemical potential $\mu$. Here, $\delta\mu=0$. There is no solution for $y(t)=0$ (i.e., $\alpha_2=0$) above some chemical potential value ($\mu \approx 1.045 T_{c}(0)$ for $\delta\mu=0$), determined by the simultaneous solution of $y(t)=0$ and $y'(t)=0$. For finite chemical potential asymmetry, $\delta\mu \ne 0$, the curves present the same behavior but with shifted minima.}
\end{figure}

The zeros of $y(t)$ for a given $\mu_{\uparrow}$ and $\mu_{\downarrow}$, i.e., for a given $\mu$ and $\delta \mu$, are the respective $T_c$. This defines the (second-order) $T_c$ versus $\mu$ phase diagram (for fixed $\delta \mu$). As can be seen in Fig.~\ref{y}, the graphical analysis of $y(t)$ shows that there is no solution for this function for $\mu$ (and $\delta \mu$) above certain values, that, as previously discussed, approach $\mu_{tc}$. Thus, within this approach, at $\mu_{tc}$ we have $y=y'=0$, where $y'=dy/dt$. The function $y(t)'$ is given by
\begin{eqnarray}
\label{y-3}
y(t)' &=&  d \left[ \frac{1}{t} - \frac{2a}{t^3} +  \frac{4b}{t^5} - \frac{6c}{t^7}\right].
\end{eqnarray}

In what follows, we discard for the moment the last term in Eqs.~(\ref{y-2}) and (\ref{y-3}) by taking $c=0$ in these equations. We will consider this term latter in the numerical analysis. These two equations, $y=0$ and $y'=0$, give $t_{tc}$ and $\mu_{tc}$ for the approximate tricritical point $P_{tc}=(\mu_{tc},t_{tc})$.  

From the equation $y(t)'=0$ we obtain $t^2=  a + \sqrt{a^2 - b^2}$. 
Self-consistently solving the above equation together with  $y(t)=0$, or
\begin{eqnarray}
\label{y-5}
\ln \left( t \right) = - \frac{a}{t^2} +  \frac{b}{t^4} = \frac{a}{a + \sqrt{a^2 - b^2}} -1,
\end{eqnarray}
we find $t = e^{ \frac{a}{a + \sqrt{a^2 - b^2}} -1 }$ and $a + \sqrt{a^2 - b^2} = \left(  e^{ \frac{a}{a + \sqrt{a^2 - b^2}} -1 } \right)^2$.

For $b=0$, i.e., at second order in the $\mu_{\uparrow,\downarrow}/T$ expansion, we obtain
\begin{equation}
a_{tc}=\frac{7 \xi(3)}{8 \pi^2} \frac{(\mu_{\uparrow}^2 + \mu_{\downarrow}^2)_{tc}}{(k_B T_c(0))^2}=
\frac{7 \xi(3)}{4 \pi^2} \frac{[\mu^2 + (\delta\mu)^2]_{tc} }{(k_B T_c(0))^2}=
\frac{1}{2e},~~~~~~~t_{tc}=\frac{T_{tc}}{T_c(0)}=\frac{1}{\sqrt{e}}.
\label{y-8}
\end{equation}

For higher-order approximations, $a_{tc}$ (and, consequently, $\mu_{tc}$) and $t_{tc}$ 
can be obtained by the same procedure, with more complex equations not shown here.
It predicts $\mu_{tc} \approx 0.527\Delta_0$ and $T_{tc} \approx 0.344\Delta_0$, in the second-order approximation, and $\mu_{tc} \approx 0.592\Delta_0$ and $T_{tc} \approx 0.321\Delta_0$, in the sixth-order approximation (which means to consider $b \neq$ and $c \neq 0$ in Eqs.~(\ref{y-2}) and (\ref{y-3})), to be compared to the aforementioned numerical result $\mu_{tc} = 0.608 \Delta_0$ and $T_{tc} = 0.318 \Delta_0$ \cite{Wolff}. Fig. \ref{mutcvsdmu} shows the behavior of the chemical potential (Fig.\ref{mutcvsdmu}(a)) and temperature (Fig.{\ref
{mutcvsdmu}(b)) at the tricritical point as a function of the chemical potential asymmetry, $\delta\mu$, 
within the expansion up to orders $(\mu_{\uparrow,\downarrow}/T)^2$ and $(\mu_{\uparrow,\downarrow}/T)^6$. The curves, in both diagrams, display the same behavior, although there are quantitative corrections. At order $\mu^2_{\uparrow,\downarrow}/T^2$, the tricritical point temperature does not vary with $\delta\mu$, while up to order $\mu^6_{\uparrow,\downarrow}/T^6$, the tricritical point temperature smoothly increases as $\delta\mu$ increases (Fig. \ref{mutcvsdmu}(b)). The chemical potential (Fig. \ref{mutcvsdmu}(a)) displays the same behavior obtained by numerical evaluation at high temperatures (compare with the $T=0.4 \Delta_0$ diagram in Fig. (\ref{FiniteT}), as we shall discuss later, for instance).

\begin{figure}[h]
\centering
\includegraphics[height=2.8in]{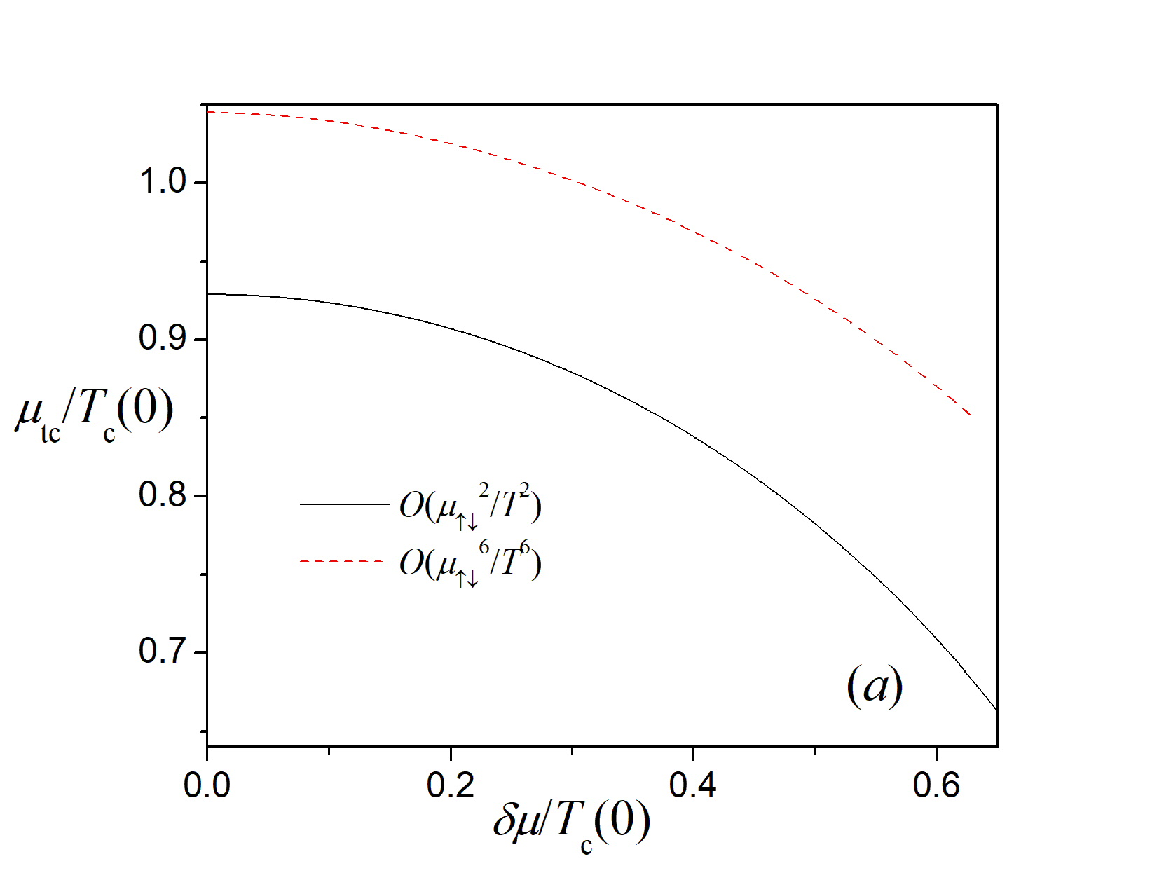}\includegraphics[height=2.8in]{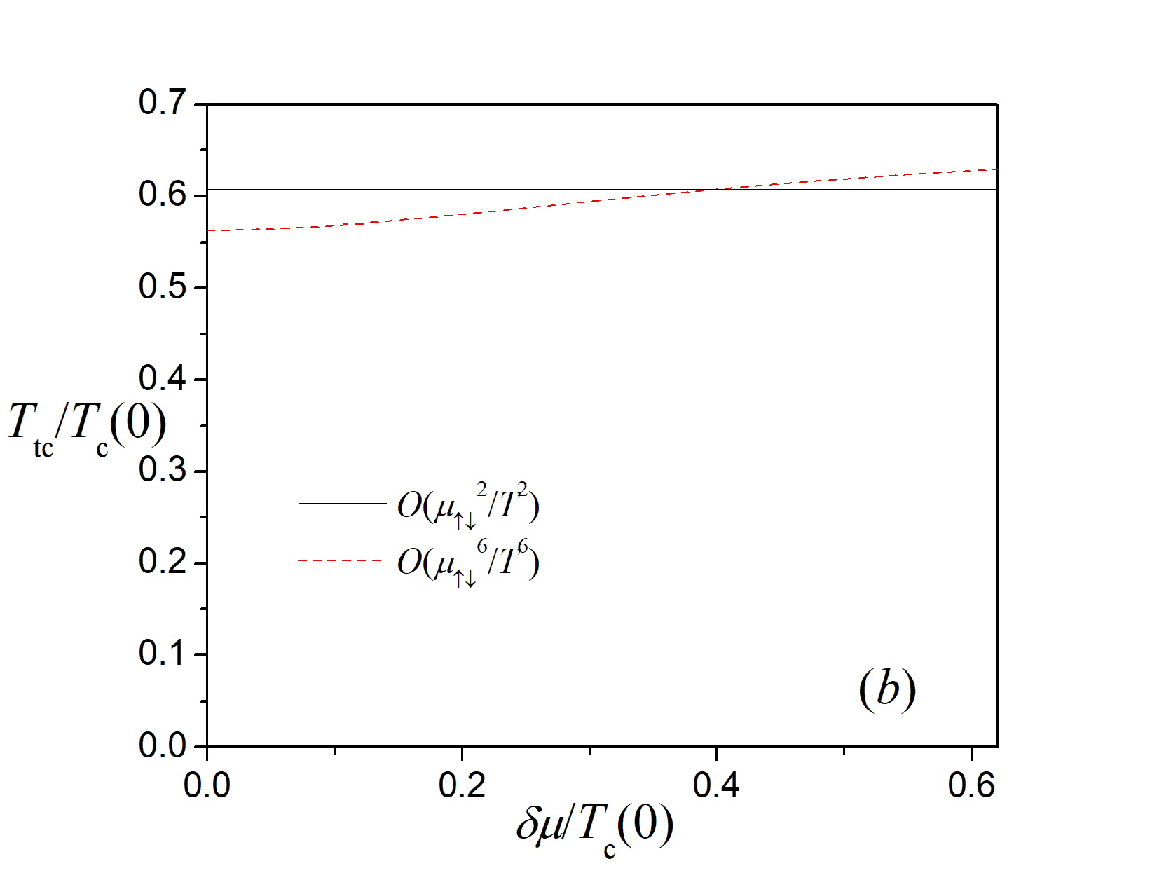}
\caption{\label{mutcvsdmu} (color online) The chemical potential (a) and temperature (b) at the tricritical point as a function of the chemical potential asymmetry $\delta\mu$ computed up to the second (solid black line) and sixth (dashed red line) order in the high-temperature expansion on $\mu_{\uparrow,\downarrow}/T$. }
\end{figure}

The number densities $n_{\uparrow,\downarrow}= -\frac{\partial}{\partial \mu_{\uparrow,\downarrow}} V_{eff}(\Delta,\mu_{\uparrow,\downarrow},T)$ read

\begin{equation}
n_{\uparrow,\downarrow}= \int^{\infty}_{0}{\frac{dp}{\pi \hbar}}~ \left[ n_k(E_{\uparrow,\downarrow}^{-}) - n_k(E_{\uparrow,\downarrow}^{+}) \right],
\label{n1}
\end{equation}
where $n_k(E_{\uparrow,\downarrow}^{+,-})=\frac{1}{e^{\beta E_{\uparrow,\downarrow}^{+,-}}+1}$ is the Fermi distribution function. The density difference

\begin{equation}
\delta n = n_{\uparrow}-n_{\downarrow}
\label{n2}
\end{equation}
is zero if $\delta \mu = \frac{g}{2} \mu_B B_0=0$, at any temperature, since in this case we have the equalities $n_k(E_{\uparrow}^{+})=n_k(E_{\downarrow}^{+})$, and $n_k(E_{\uparrow}^{-})=n_k(E_{\downarrow}^{-})$. The physical meaning of these results is that at zero external Zeeman magnetic field, the $\uparrow$ (up) and $\downarrow$ (down) electrons of the conduction $(+)$ band have the same density, and the same for the electrons of the valence $(-)$ band.

In the high temperature regime, the number densities are given by 

\begin{equation}
n_{\uparrow,\downarrow}(T)= \frac{1}{ \pi \hbar v_F} \left[ 1 - \frac{7 \zeta(3)}{4 \pi^2} \frac{\Delta^2}{(k_B T)^2} \right]\mu_{\uparrow,\downarrow}.
\label{nd1}
\end{equation}
In the high temperature limit the total number density, $n_{T}(T)=n_{\uparrow}(T)+n_{\downarrow}(T)$, is independent of the applied field, as happens at zero $T$,

\begin{equation}
n_{T}(T)= \frac{2}{ \pi \hbar v_F} \left[ 1 - \frac{7 \zeta(3)}{4 \pi^2} \frac{\Delta^2}{(k_B T)^2} \right]\bar \mu,
\label{nT}
\end{equation}
and for the density difference we obtain

\begin{equation}
\delta n_{high~T}(T, \delta \mu)= \frac{2}{ \pi \hbar v_F} \left[ 1 - \frac{7 \zeta(3)}{4 \pi^2} \frac{\Delta^2}{(k_B T)^2} \right]\delta \mu,
\label{nd2}
\end{equation}
that, as we have observed before, is clearly zero if $B_0=\delta \mu = 0$. Since the densities have to be evaluated at the minimum of the effective potential, we use $\Delta^2$ from Eq.~(\ref{poteff6}) in the equation above and find the temperatures at which the densities and, consequently, the density difference vanish. These temperatures are the solutions to

\begin{equation}
-\frac{1}{2} + \ln \left( \frac{e^{\gamma_E }\Delta_0}{\pi k_B T_c^*} \right) + \left(\frac{186 \zeta(5)}{112 \pi^2 \zeta(3)} - \frac{7 \zeta(3)}{8 \pi^2} \right)  \frac{(\mu_{\uparrow}^2+\mu_{\downarrow}^2)}{(k_B T_c^*)^2} =0,
\label{nd3}
\end{equation}
where $T_c^*=T_c^*(\mu_{\uparrow},\mu_{\downarrow})$. As for the gap parameter, the equation above defines the second-order line for the densities and the density imbalance. At $\mu_{\uparrow}=\mu_{\downarrow}=0$, we get

\begin{equation}
T_c^*(\mu_{\uparrow}=\mu_{\downarrow}=0) \equiv T_c^*(0)=\frac{e^{\gamma_E-\frac{1}{2}}}{\pi} \frac{\Delta_0}{k_B}.
\label{Tc*1}
\end{equation}
It is very easy to see that $T_c^*(0)=\frac{T_c(0)}{\sqrt{e}}\approx 0.6 T_c(0)$, where $T_c(0)$ is given by Eq.~(\ref{Tc2}). This result shows that the densities vanish before the gap goes to zero.

From Fig.~\ref{mutcvsdmu}, we observe that, in the high-temperature regime, as the chemical potential difference increases (i.e., as the external magnetic field increases) the critical chemical potential at any fixed temperature decreases. This behavior contrasts with the zero temperature result depicted in Fig. \ref{zeroTpd}. In this case, the critical chemical potential decreases as $\delta \mu$ increases up to $\delta \mu \approx 0.249 \Delta_0$. At this point, the critical line acquires a positive slope, and the critical chemical potential grows with $\delta \mu$ up to high values of $\delta \mu / \Delta_0$. This zero temperature behavior of the critical curve allows the reentrant (gapped-gapless-gapped) phase transition, as observed in subsection III.2. It is interesting to investigate how the system evolves between these two different regimes. Obviously, it cannot be obtained from the high temperature expansion we obtained in Eq. (\ref{Veff}), but from the direct (numerical) evaluation of Eq. (\ref{poteff4}). This procedure is a little time-consuming, since one has to go through several different points in the $\mu/\Delta_0 \times \delta \mu/\Delta_0$ diagram to determine the critical curves at different temperatures. The result is shown in Fig. \ref{FiniteT}, which brings the critical curves in the  $\mu/\Delta_0 \times \delta \mu/\Delta_0$ plane at different temperatures. As one can see, for low temperatures (e.g., $T=0.1 \Delta_0$) the behavior of the critical line is similar to the $T=0$ case. For intermediate $T$ values ($T/\Delta_0 = 0.2, 0.25, 0.3$), the increasing of the magnetic field (increasing $\delta \mu$) still leads to a first-order phase transition from a gapless to a gapped phase at some finite $\delta \mu$ value. However, as $\delta \mu$ increases to higher values approaching $\delta \mu / \mu = 1$, the critical chemical potential decreases.
Above $T \approx 0.32\Delta_0$, the reentrant phase transition is no longer present, and the chemical potential at the tricritical point monotonically decreases as $\delta \mu$ increases, showing the same behavior we found in the high temperature limit (see Fig. \ref{mutcvsdmu}, for instance). In Fig. \ref{FiniteT}, the critical lines are limited by $\delta \mu \le \mu$, as indicated by the straight line $\delta \mu = \mu$.

\begin{figure}[h]
\centering
\includegraphics[height=4.0in]{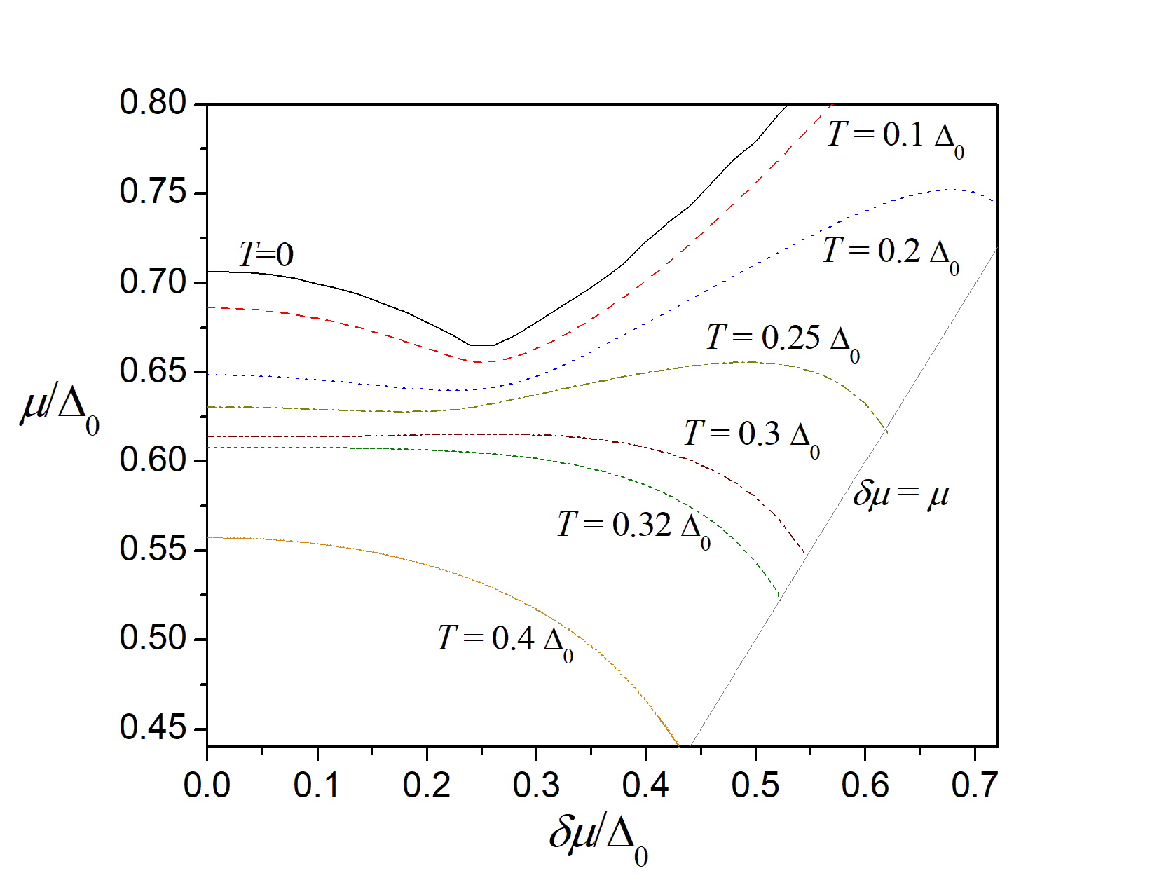}
\caption{\label{FiniteT}
(color online) Finite temperature critical lines in the $\mu/\Delta_0$ versus $\delta \mu/\Delta_0$ plane. For low $T$, the behavior is similar to that of the $T=0$ case, with a reentrant (gapped-gapless-gapped) phase transition for $\mu$ slightly below the critical potential at $\delta \mu = 0$. At $T \approx 0.32 \Delta_0$ and above, the reentrant phase is no longer observed, and the behavior of the critical line coincides with the one obtained in the high-temperature limit. The critical lines end at $\delta \mu = \mu$, represented by the straight line on the diagram.}
\end{figure}

\subsection{Magnetic Properties}
\label{magprop}

The Pauli magnetization of the system in the high temperature limit has the following expression:

\begin{eqnarray}
M_{high~T}(T) &=& \mu_B \delta n_{high~T}(T) = \frac{2 \mu_B}{ \pi \hbar v_F} \left[ 1 - \frac{7 \zeta(3)}{4 \pi^2} \frac{\Delta^2}{(k_B T)^2} \right]\delta \mu \nonumber \\
&=& \frac{g \mu_B^2}{ \pi \hbar v_F} \left[ 1 - \frac{7 \zeta(3)}{4 \pi^2} \frac{\Delta^2}{(k_B T)^2} \right] B_0,
\label{mag1}
\end{eqnarray}
where $\Delta^2$ is given Eq.~(\ref{poteff6}),
\begin{equation}
\Delta^2 = -\frac{\alpha_2}{2 \alpha_4} = - \frac{8 \pi^2}{7 \zeta(3)} (k_B T)^2 \left[ \ln \left( \frac{\pi k_B T}{e^{\gamma_E} \Delta_0} \right) + \frac{7 \zeta(3)}{8 \pi^2} \frac{(\mu_{\uparrow}^2+\mu_{\downarrow}^2)}{(k_B T)^2} \right]
\left( 1 + \frac{186 \zeta(5)}{56 \zeta(3) \pi^2} \frac{(\mu_{\uparrow}^2+\mu_{\downarrow}^2)}{(k_B T)^2} \right).
\label{gapT}
\end{equation}

The second-order line where the magnetization vanishes is the same as the one given by Eq.~(\ref{nd3}). Finally, we obtain the magnetic susceptibility in this regime

\begin{equation}
\chi_{high~T}(T)= \frac{\partial M_{high~T}(T)}{\partial B_0}= \chi(0)+ \chi(T),
\label{ms1}
\end{equation}
where
\begin{equation}
\chi(0)=\frac{g \mu_B^2}{ \pi \hbar v_F},
\label{ms2}
\end{equation}
and

\begin{equation}
\chi(T)=-\frac{g \mu_B^2}{ \pi \hbar v_F} \left[\frac{7 \zeta(3)}{4 \pi^2} \frac{\Delta^2}{(k_B T)^2} \right],
\label{ms3}
\end{equation}
where $\Delta^2$ is given Eq.~(\ref{gapT}).

$\chi(0)$ is the well-known zero temperature contribution for the Pauli expression of the magnetic susceptibility for noninteracting electrons. The function $\chi_{high~T}(T)$ also behaves at finite temperature as the densities and the magnetization, with a second-order transition up to the tricritical point given by Eq.~(\ref{nd3}). Below this point, the transition is again of first-order. 

Notice that in spite of the full polarization, at $B_{0,c}$ the magnetization is given by a similar expression to that shown in Eq.~(\ref{mag1}). With the help of Eq.~(\ref{nd1}), we find:

\begin{eqnarray}
M_{high~T,c}(T)^{fp} = \mu_B n_{\uparrow}(T) &=&  \frac{\mu_B}{ \pi \hbar v_F} \left[ 1 - \frac{7 \zeta(3)}{4 \pi^2} \frac{\Delta^2}{(k_B T)^2} \right](\mu_c + \delta \mu_c)
\nonumber \\
&=&\frac{\mu_B}{ \pi \hbar v_F} \left[ 1 - \frac{7 \zeta(3)}{4 \pi^2} \frac{\Delta^2}{(k_B T)^2} \right](\mu_c + \frac{g}{2}\mu_B B_{0,c}), 
\label{mag2}
\end{eqnarray}
where {\it fp} stands for fully polarized and $\Delta^2$ is also given by Eq.~(\ref{gapT}). We would like to point out that the increase in the $g$ factor can, again (as discussed below Eq.~(\ref{cmf2})), play an important role in this system; here, in the increase in both magnetization and magnetic susceptibility.

\section{Inhomogeneous $\Delta(x)$ Condensates}
\label{IC}

In this section we consider inhomogeneous $\Delta(x)$ condensates as solutions of the gap equation in Eq.~(\ref{action4}). As has been pointed out several times in the literature, the assumed spatially homogeneous condensate does not allow the emergence of a crystalline phase~\cite{gnpolymers} and its subsequent investigation. Besides, since we consider the introduction of a chemical potential in the theory, and the effects of an external Zeeman magnetic field on the system, an important fact should be taken into account. It has been shown that doping in conducting polymers with degenerate ground states results in lattice deformation, or non-linear excitations, such as kink solitons and polarons, which means that $\Delta(x)$ can vary in space~\cite{Horovitz1,Horovitz2,Review}. Therefore, one may expect not only homogeneous configurations, as considered in the previous sections, but also that the inclusion of these excitations in theoretical calculations in this model should not be disregarded. Within the GN field theory model we are considering here, taking into account kink-like configurations in the large $N$ approximation, the authors of Refs.~\cite{gnpolymers,gnpolymers1,gnpolymers2,gnpolymers3,gnpolymers4,Basar} found evidence for a crystalline phase that shows up in the extreme $T \sim 0$ and large $\mu$ part of the phase diagram.
However, in the other extreme of the phase diagram (large $T$ and small $\mu$), results for the critical temperature and tricritical point remain unaltered when compared to those obtained from the standard large $N$ results, as the ones from the early work on the GN model~\cite{Wolff}. 

\subsubsection{GL expansion to ${\cal O}(\alpha_{4})$ for the GN model}

To take into account the effects of inhomogeneous configurations in the GL expansion of the grand potential density, let us write Eq.~(\ref{Veff}) in terms of $\Delta(x)$ and its derivatives up to $\alpha_4$~\cite{gnpolymers,gnpolymers1,gnpolymers2,gnpolymers3,gnpolymers4,Basar}:
\begin{equation}
\label{Veff(x)}
V_{eff}(x)= \alpha_0 + \alpha_2 \Delta(x)^2 + \alpha_4 ( \Delta(x)^4 + \Delta^{\prime}(x)^2 ),
\end{equation}
where $\Delta^{\prime}(x) \equiv \frac{d\Delta(x)}{dx}$. Minimizing the free energy $E = \int V_{eff}(x)~dx$ as a functional of $\Delta(x)$, we get

\begin{equation}
\label{difeq1}
\Delta^{\prime\prime}_x -2 \Delta_x^3 - \frac{\alpha_2}{\alpha_4}\Delta_x=0.
\end{equation}
with $\Delta_x \equiv \Delta(x)$. Eq.~(\ref{difeq1}) has the form of a nonlinear Scr{\"o}dinger equation (NLSE). An equation of the form

\begin{equation}
\label{difeq2}
\Delta_x'' -2 \Delta_x^3 + (1+\nu) \Delta_0^2 \Delta_x=0,
\end{equation}
has a general solution, parametrized by $\Delta_0$ and $\nu$, given by~\cite{Basar}

\begin{equation}
\label{difeq3}
\Delta_x= \Delta_0 \sqrt{\nu} {\rm sn} (\Delta_0 x; \nu),
\end{equation}
where ${\rm sn}$ is the Jacobi elliptic function with the real elliptic parameter $0 \leq \nu \leq  1$. The ${\rm sn}$ function has period $2 {\bf K}(\nu)$, with ${\bf K}(\nu) \equiv \int_0^{\pi/2}[1-\nu \sin^2(t)]^{-1/2} dt$, the complete elliptic integral of first kind. The gap (mass) modulation function $\Delta_x$ in Eq.~(\ref{difeq3}) represents a set of real kinks, reducing to the single kink condensate given in Eq.~(\ref{kink}) when $\nu=1$. Comparing Eqs.~(\ref{difeq1}) and (\ref{difeq2}) one can identify the scale parameter $\Delta_0$ as

\begin{equation}
\label{difeq4}
\Delta_0=\sqrt{\left( -\frac{\alpha_2}{\alpha_4} \right) \left( \frac{1}{1+\nu} \right)}.
\end{equation}
Since $\frac{1}{1+\nu} > 0$, the solution for inhomogeneous condensates only has physical meaning if  $\frac{\alpha_2}{\alpha_4} < 0$, as in the case of homogeneous condensates.

The term $- \frac{\alpha_4}{3} (\Delta_x^2)^{\prime\prime}$ can be added to $V_{eff}(x)$ in Eq.~(\ref{Veff(x)}) without altering the solution for the minimum of the free energy, Eq.~(\ref{difeq1}) \cite{Basar}. By using the following identities, supported by the solution in (\ref{difeq3}):

\begin{equation}
\label{difeq5}
(\Delta^{\prime}_x)^2 = \Delta_x^4 - (1+\nu) \Delta_0^2 \Delta_x^2 + \nu \Delta_0^4,
\end{equation}
and

\begin{equation}
(\Delta_x^2)^{\prime\prime} = 6\Delta_x^4 - 4(1+\nu) \Delta_0^2 \Delta_x^2 + 2 \nu \Delta_0^4,
\end{equation}
we can rewrite the $x$-dependent grand potential density as

\begin{equation}
\label{Veff(x)2}
V_{eff}(x)= \alpha_0 + \alpha_2\Delta_x^2 +  \frac{\alpha_4}{3} \left( (1+\nu)\Delta_0^2 \Delta_x^2 + \nu \Delta_0^4 \right).
\end{equation}

To compare this result with the homogeneous case, one can average Eq.(\ref{Veff(x)2}) over one entire period and, since $\langle \Delta_x^2 \rangle$ can be explicitely computed, resulting in $\langle \Delta_x^2 \rangle = \left(1-\frac{{\bf E}(\nu)}{{\bf K}(\nu)} \right) \Delta_0^2 $, we obtain the effective grand potential of this {\it crystal phase} as
\begin{equation}
\label{Veff(x)3}
V_{eff}^{crystal} \equiv \langle V_{eff}^{crystal}(x) \rangle = \alpha_0 +  {\cal A}_2 \Delta_0^2  + {\cal A}_4 \Delta_0^4.
\end{equation}
Here, ${\bf E}(\nu)$ is the complete elliptic integral of second kind and ${\bf K}(\nu)$ is, as before, the complete elliptic integral of the first kind. The ratio ${\bf E}(\nu)/{\bf K}(\nu)$ is a smooth function of $\nu$ interpolating monotonically between $0$ and $1$. The coefficients of the GL expansion are
\begin{eqnarray}
\label{Newcoef}
{\cal A}_2 &=& \alpha_2 \left(1-\frac{{\bf E}(\nu)}{{\bf K}(\nu)} \right),\\
\nonumber
{\cal A}_4 &=&  \frac{\alpha_4}{3} \left[\nu + (1+ \nu)\left(1-\frac{{\bf E}(\nu)}{{\bf K}(\nu)} \right)\right].
\end{eqnarray}

This result generalizes the analysis of the grand potential density (up to order $\alpha^4$), and the results obtained in Section~\ref{rep} for homogeneous condensates are obtained as particular cases when specific values of the elliptic parameter $\nu$ are taken. Namely,
\newline
\newline 
(${\bf 1.}$) For $\nu=0$, $\frac{{\bf E}(\nu=0)}{{\bf K}(\nu=0)}=1$, so the grand potential is that of the gapless (massless) phase, for which $\Delta_0=0$ and $V_{eff}^{crystal}(\nu=0)=\alpha_0$.
\newline
\newline
(${\bf 2.}$) For $\nu=1$, $\frac{{\bf E}(\nu=1)}{{\bf K}(\nu=1)}=0$, such that $V_{eff}^{crystal}(\nu=1)=\alpha_0 +  {\alpha}_2 \Delta_0^2  + {\alpha}_4 \Delta_0^4=\alpha_0 -\frac{\alpha_2}{4\alpha_4}=V_{eff}$, where $V_{eff}$ is the grand potential of the homogeneous case at the non-trivial minimum $\Delta_0$, Eq.~(\ref{Veff}), up to order $\alpha_4$. 
\newline
\newline
(${\bf 3.}$) for $0< \nu < 1$, the grand potential is that of Eq.~(\ref{Veff(x)3}) that, as expected, is the interpolation of the two previous cases. It is of particular interest for our results, however, to notice, from Eq.~(\ref{Veff(x)3}), that for any $0 \leq   \nu \leq  1$, the tricritical point is still found for $\alpha_2=\alpha_4=0$. These coefficients, which are $\mu_{\uparrow,\downarrow}$ and $T$ dependent, are not affected by the space dependence of the condensate $\Delta(x)$.

\subsubsection{GL expansion to ${\cal O}(\alpha_{6})$ for the GN model}

In order to go to ${\cal O}(\alpha_{6})$ in the GL expansion, the grand potential density is expanded in powers of the real condensate field $\Delta_x$ and its derivatives (the total derivative terms are dropped, since these terms are not relevant here)~\cite{Basar}:

\begin{equation}
\label{Veff(x)-2}
V_{eff}(x)= \alpha_0 + \alpha_2 \Delta_x^2 + \alpha_4 ( \Delta_x^4 + {\Delta_x^{\prime}}^2 ) + \tilde\alpha_6 (2 \Delta_x^6 + 10 \Delta_x^2 {\Delta_x^{\prime}}^2 + {\Delta_x^{\prime\prime}}^2),
\end{equation}
where for convenience we have defined $\tilde \alpha_6 $ in the above equation in terms of the one in Eq.~(\ref{Veff}) as $\tilde \alpha_6 = \alpha_6/2$. From Eq.~(\ref{Veff(x)-2}) one obtains the following GL equation

\begin{equation}
\label{GL-1}
\Delta_x^{\prime\prime\prime\prime} - 10 \Delta_x^2  \Delta_x^{\prime\prime} - 10 \Delta_x  (\Delta_x^{\prime})^2 + 6 \Delta_x^5 + \frac{\alpha_4}{\tilde \alpha_6} ( -\Delta_x^{\prime\prime} +2\Delta_x^3) + \frac{\alpha_2}{\tilde \alpha_6} \Delta_x=0.
\end{equation}
For a homogeneous condensate $\Delta_x=\Delta_0$ the above equation reduces to

\begin{equation}
\label{GL-2}
6 \Delta_0^5 + 2 \frac{\alpha_4}{\tilde \alpha_6} \Delta_0^3 + \frac{\alpha_2}{\tilde \alpha_6} \Delta_0=0,
\end{equation}
with the trivial solution

\begin{equation}
\label{GL-3}
\Delta_0=0,
\end{equation}
corresponding to a gapless homogeneous phase, and

\begin{equation}
\label{GL-4}
\Delta_{0 \pm}^2= - \frac{\alpha_4}{6\tilde \alpha_6}\left( 1 \pm \sqrt{1-\frac{6\alpha_2 \tilde \alpha_6}{\alpha_4^2}} \right),
\end{equation}
representing a gapped homogeneous phase, agreeing with Eq.~(\ref{Deltamin}). A spatially dependent solution to Eq.~(\ref{GL-1}) is very difficult to obtain. However, the inhomogeneous solution to the NLSE, $\Delta_x= \Delta_0 \sqrt{\nu} {\rm sn} (\Delta_0 x; \nu)$ (Eq.~(\ref{difeq3})), can be used~\cite{Basar}. $\Delta_x(\Delta_0,\nu)$ is also solution of the nonlinear equations:

\begin{eqnarray}
\label{GL-5}
-\Delta_x^{\prime\prime} + 2 \Delta_x^3 -(1+\nu) \Delta_0^2 \Delta_x = 0,\\
\nonumber
\Delta_x^{\prime\prime\prime\prime} - 10 \Delta_x^2  \Delta_x^{\prime\prime} - 10 \Delta_x  (\Delta_x^{\prime})^2 + 6 \Delta_x^5 - (\nu^2 + 4\nu + 1) \Delta_0^4 \Delta_x=0.
\end{eqnarray}
Then, with the help of these two equations, Eq.~(\ref{GL-1}) reduces to

\begin{equation}
\label{GL-6}
\Delta_0^4 + \frac{(\nu+1)}{(\nu^2+ 4\nu +1)} \frac{\alpha_4}{\tilde \alpha_6} \Delta_0^2 + \frac{1}{(\nu^2+ 4\nu +1)} \frac{\alpha_2}{\tilde \alpha_6}=0,
\end{equation}
which has as solutions

\begin{equation}
\label{GL-7}
\Delta_{0 \pm}^2 = -\frac{b}{2} \left[ 1\pm \sqrt{1 - \frac{4}{b(\nu+1)} \frac{\alpha_2}{\alpha_4}} \right],
\end{equation}
where $b \equiv \frac{(\nu+1)}{(\nu^2+ 4\nu +1)} \frac{\alpha_4}{\tilde \alpha_6}$.
Then, the grand potential evaluated on the crystalline solution up to ${\cal O}(\alpha_{6})$ is 

\begin{equation}
\label{GL-8}
V_{eff}^{crystal} = \alpha_0 +  {\cal A}_2 \Delta_0^2  + {\cal A}_4 \Delta_0^4 + {\cal A}_6 \Delta_0^6,
\end{equation}
where $\Delta_0$ is given by the solution in Eq.~(\ref{GL-7}) which yields the lowest value of the grand potential density, ${\cal A}_2$ and ${\cal A}_4$ are the same as in Eq.~(\ref{Newcoef}), and

\begin{equation}
\label{GL-9}
{\cal A}_6 =  \frac{\tilde \alpha_6}{5} \left[2\nu(\nu+1) + (\nu^2 + 4\nu + 1)\left(1-\frac{{\bf E}(\nu)}{{\bf K}(\nu)} \right)\right].
\end{equation}

The above results generalize the analysis of the grand potential density in the GL expansion to ${\cal O}(\alpha_{4})$, such that the homogeneous case is obtained as a particular case when specific values of the elliptic parameter $\nu$ are chosen:
\newline
\newline 
(${\bf 1.}$) For $\nu=0$, $\frac{{\bf E}(\nu=0)}{{\bf K}(\nu=0)}=1$, so the grand potential is that of the gapless (massless) phase, for which $\Delta_0=0$ and ${\cal A}_2 = {\cal A}_4 = {\cal A}_6 =0$, resulting in $V_{eff}^{crystal}(\nu=0)=\alpha_0$.
\newline
\newline
(${\bf 2.}$) For $\nu=1$, $\frac{{\bf E}(\nu=1)}{{\bf K}(\nu=1)}=0$, such that $V_{eff}^{crystal}(\nu=1)=\alpha_0 +  {\alpha}_2 \Delta_0^2  + {\alpha}_4 \Delta_0^4 + {\alpha}_6 \Delta_0^6 =V_{eff}$, where $V_{eff}$ is the grand potential of the homogeneous case, Eq.~(\ref{Veff}), at the non-trivial minimum $\Delta_0$. 
\newline
\newline
(${\bf 3.}$) for $0< \nu < 1$, the grand potential is the one in Eq.~(\ref{GL-8}) that, as expected, is the interpolation of the two previous cases.  

Although to ${\cal O}(\alpha_{6})$ of the GL expansion a crystal phase appears in a tiny region below the tricritical point in the $(T,\mu)$ phase diagram~\cite{Basar}, the tricritical point itself, found for ${\cal A}_2={\cal A}_4=0$, which implies $\alpha_2=\alpha_4=0$ for any $0 \leq   \nu \leq  1$ is, therefore, the same as in the homogeneous case.

Then, considering inhomogeneous $\Delta(x)$ condensates (and consequently a $x$-dependent grand potential density) does not alter the location of the tricritical point of the GN phase diagram under the influence of a given (fixed) external Zeeman magnetic field. However, it is worth remarking that the actuation of an increasing external Zeeman field is sufficient to change the position of the tricritical point reducing, in this way, the gapped region in the phase diagram as the external magnetic field increases, as we have seen before (see, for instance, figures~\ref{distinct} and~\ref{mutcvsdmu}).

\section{Conclusions}
\label{Summary}

To summarize, we have studied the mean-field phase diagram of 1D systems that can be modeled with a GN-type model under an external Zeeman magnetic field $B_0$. The system can be converted into partially or fully polarized, depending on the intensity of the $B_0$. Since the GN model can be employed as an effective model for {\it trans}-polymers, the investigation made in this work can be relevant in studies of 1D systems of interest in condensed matter, such as, for instance, semiconductor devices, as diodes and transistors~\cite{Diode}, and light-emitting diodes~\cite{Led}, based on conjugated polymers.

After the employment of a Ginzburg-Landau expansion of the effective potential, up to orders $(\Delta/k_BT)^6$ and $(\mu_{\uparrow,\downarrow}/k_BT)^4$, the phase diagrams in the plane $(\mu_{\uparrow,\downarrow},T)$ were obtained. This allowed the identification of the first and second-order lines, as well as the location of the tricritical point in the $t$ $\left(= T/T_c(0) \right)$ versus $\mu/T_c(0)$ phase diagram for various strengths of the external Zeeman magnetic field, generalizing previous investigations at $\delta \mu = B_0 = 0$.

At zero temperature, we have found that departing from the system in the gapless (massless) phase, i.e., with $\mu=\mu_c$ and $\delta \mu=0$, there is a first-order phase transition to a gapped phase as the magnetic field is increased from zero and reaches the critical value $\delta \mu_c=0.372 \Delta_0$. This suppression of the gapless phase can be seen as the 1D analogous to the one found in a 2D system, in which the Zeeman magnetic field suppresses the metallic behavior, inducing a metal-insulator transition at a critical field strength~\cite{Dent}. If, instead, we start with some $\mu < \mu_c$ (i.e., in a gapped phase) and turn on the Zeeman field, as $\delta \mu$ is increased, there is a ``reentrant'' behavior, with gapped-gapless-gapped phases, see Fig.~\ref{zeroTpd}. The gapless-gapped phase transition at a critical field strength $B_{0,c}$ as a function of temperature was analyzed, and we found that the reentrant phenomenon was observed only for low temperatures, or $T\lesssim 0.3 \Delta_0$.

We also investigated the GN phase diagram under the influence of an external Zeeman magnetic field $B_0$ considering spatially inhomogeneous $\Delta(x)$ condensates. We found that up to order $\alpha_6$ of the GL expansion the position of the tricritical point in the phase diagram is not altered. Still in the context of spatially inhomogeneous $\Delta(x)$ condensates, it would be interesting to study the influence of the Zeeman magnetic field in the tricritical point of the $(T,\mu)$ plane, considering the exact crystalline solutions to the associated gap equation~\cite{Basar}.

A direct extension of the present work would be the investigation of the metal-insulator transition in 1D doped systems under the influence of a Zeeman magnetic field, as well as the spin-polarized phases and the magnetic properties of 1D {\it trans}- and {\it cis}-polymers. In the former situation, one considers the massless Gross-Neveu model, as we studied here, while in the latter case, one has to take into account the massive GN model~\cite{Aragao}.

As a final remark, it would be interesting to study the possibility of using these 1D systems as spin-polarized conductors upon the joint application of a Zeeman magnetic field and an electric field in the system. We intend to address this topic in the future.\\
\\
We would like to thank M. Continentino for valuable conversations. We are grateful to M. Thies for enlightening and helpful discussions. H.C. acknowledges partial support by FAPEMIG. 
\\
\\


\begin{thebibliography}{99}

\bibitem{GN} D. Gross and A. Neveu, Phys. Rev. D {\bf 10}, 3235 (1974).

\bibitem{Coleman} S. Coleman, {\it Aspects of Symmetry}, Cambridge University Press, (1988).

\bibitem{Klimenko1} D. Ebert, K. G. Klimenko, A. V. Tyukov, and V. Ch. Zhukovsky, Phys. Rev. D {\bf 78}, 045008 (2008).

\bibitem{Klimenko2} V. Ch. Zhukovsky, K. G. Klimenko, V. V. Khudyakov, and D. Ebert, JETP Lett. {\bf 73}, 121 (2001).

\bibitem {Klimenko3} K. G. Klimenko and R. N. Zhokhov, Phys. Rev. D {\bf 88}, 105015 (2013).

\bibitem{Lenz1} J. J. Lenz, M. Mandl, and A. Wipf, Phys. Rev. D {\bf 107}, 094505 (2023).

\bibitem{CM} A. Chodos and H. Minakata, Phys. Lett. {\bf A191}, 39 (1994); Nucl. Phys. {\bf B490}, 687 (1997).

\bibitem{CaldasJSTAT} H. Caldas, J. Stat. Mech. {\bf 2011}, P10005 (2011).

\bibitem{Thies1} M. Thies, Phys. Rev. D {\bf 102}, 096006 (2020).

\bibitem{Lenz2} J. J. Lenz, L. Pannullo, M. Wagner, B. Wellegehausen, and A. Wipf, Phys. Rev. D {\bf 101}, 094512 (2020).

\bibitem{Lenz3} J. J. Lenz, M. Mandl, and A. Wipf, Phys. Rev. D {\bf 105}, 034512 (2022).

\bibitem{Lobos} L. M. Arancibia, C. G. S\'anchez, and A. M. Lobos, Phys. Rev. B {\bf 106}, 245426 (2022).

\bibitem{Nei} N. Lopes, M. A. Continentino, and D. G. Barci, Phys. Rev. B {\bf 105}, 165125 (2022).

\bibitem{Thies2} M. Thies, Phys. Rev. D {\bf 105}, 116003 (2022).

\bibitem{Thies3} M. Thies, Phys. Rev. D {\bf 106}, 056026 (2022).

\bibitem{Mao} A. Chodos, F. Cooper, W. Mao, H. Minakata, and A. Singh, Phys. Rev. D {\bf 61}, 045011 (2000).

\bibitem{Wolff} U. Wolff, Phys. Lett. {\bf B157}, 303 (1985).

\bibitem{Chodos} A. Chodos, F. Cooper, W. Mao, A. Singh, Phys. Rev. D {\bf 63}, 096010 (2001).

\bibitem{TCP} A tricritical point is a point in the phase diagram of several systems at which the order of the transition changes its character from (discontinuous) first to (continuous) second-order.

\bibitem{Caldas2} H. Caldas, Nucl. Phys. {\bf B807}, 651 (2009).

\bibitem{Dent} P. J. H. Denteneer and R. T. Scalettar, Phys. Rev. Lett. {\bf 90}, 246401 (2003).

\bibitem{PRB2} H. Caldas and R. O. Ramos, Phys. Rev. B {\bf 80}, 115428 (2009).

\bibitem{PRB} H. Caldas, J.-L. Kneur, M. B. Pinto and R. O. Ramos, Phys. Rev. B {\bf 77}, 205109 (2008).

\bibitem{Kapusta} J. I. Kapusta, {\it Finte-temperature Field Theory}. 1 ed., New York, Cambridge University Press, (1989).

\bibitem{Kittel} C. Kittel, {\it Introduction to Solid State Physics}, sixth ed., John Wiley \& Sons, New York, (1986).

\bibitem{Madelung} O. Madelung, {\it Introduction to Solid State Theory}, 2a ed., Springer-Verlag, Berlim (1981).

\bibitem{Dashen} R. F. Dashen, B. Hasslacher and A. Neveu, Phys. Rev. D {\bf 12}, 2443 (1975).

\bibitem{Feinberg} J. Feinberg, Ann. Phys. {\bf 309}, 166 (2004), and references therein.

\bibitem{gnpolymers} M. Thies and K. Urlichs, Phys. Rev. D {\bf 67}, 125015 (2003).

\bibitem{gnpolymers1} O. Schnetz, M. Thies and K. Urlichs, Ann. Phys. (NY) {\bf 314}, 425 (2004). 

\bibitem{gnpolymers2} M. Thies and K. Urlichs, Phys. Rev. D {\bf 72}, 105008 (2005). 

\bibitem{gnpolymers3} M. Thies, J. Phys. A {\bf 39}, 12707 (2006).

\bibitem{gnpolymers4} G. Basar, and G. V. Dunne, Phys. Rev. D {\bf 78}, 065022 (2008).

\bibitem{Basar} G. Basar, G. v. Dunne, and M. Thies, Phys. Rev. D {\bf 79}, 105012 (2009).

\bibitem{SSH} W. P. Su, J. R. Schrieffer and A. J. Heeger, Phys. Rev. Lett. {\bf 42}, 1698 (1979); Phys. Rev. B {\bf 22}, 2099 (1980).

\bibitem{Review} A. J. Heeger, S. Kivelson, J. R. Schrieffer and W. P. Su, Rev. Mod. Phys. {\bf 60}, 781 (1988).

\bibitem{Fernando} J. Chen, T. -C. Chung, F. Moraes and A. J. Heegerr, Solid State Commun. {\bf 53}, 757 (1985); F. Moraes, J. Chen, T. -C.  Chung and A. J. Heeger, Synth. Met. {\bf 11}, 271 (1985).

\bibitem{CastroNeto} V. N. Kotov, B. Uchoa, V. M. Pereira, F. Guinea, and A. H. Castro Neto, Rev. Mod. Phys. {\bf 84}, 1067 (2012).

\bibitem{Xuan} F. Xuan and S. Y. Quek, npj Comput. Mater. {\bf 7}, 198 (2021).

\bibitem{Rudnei} J.-L. Kneur, M. B. Pinto, R. O. Ramos, Phys. Rev. D {\bf 74} 125020 (2006), and references therein.

\bibitem{Jacobs} L. Jacobs, Phys. Rev. D {\bf 10}, 3956 (1974); B. Harrington and A. Yildiz, Phys. Rev. D {\bf 11}, 779 (1975).

\bibitem{jstat} H. Caldas and A. L. Mota, J. Stat. Mech. {\bf 2008}, P08013 (2008).

\bibitem{Horovitz1} B. Horovitz, Solid State Comm. {\bf 34}, 61 (1980).

\bibitem{Horovitz2} J. A. Krumhansl, B. Horovitz, A. J. Heeger, Solid State Comm. {\bf 34}, 945 (1980).

\bibitem{Diode} J. H. Burroughes, C. A. Jones, and R. H. Friend, Nature {\bf 335}, 137 (1988).

\bibitem{Led} J. H. Burroughes, D. D. C. Bradley, A. R. Brown, R. N. Marks, K. Mackay, R. H. Friend, P. L. Burns and A. B. Holmes, Nature {\bf 347}, 539 (1990).

\bibitem{Aragao} C. A. A. de Carvalho, Mod. Phys. Lett. B {\bf 3}, 125 (1988).



\end{thebibliography}
\end{document}